\documentclass{aastex}
\usepackage{spr-astr-addons}
\usepackage{url}

\newcommand{\emaila}{bzhdai@ynu.edu.cn}

\begin{document}

\title{The role of jet properties in the spectral evolution of the powerful blazars}
\slugcomment{Not to appear in Nonlearned J., 45.}
\shorttitle{Short article title}
\shortauthors{Autors et al.}

\author{Wen Hu\altaffilmark{1,3}} \and \author{Wei Zeng\altaffilmark{2,3}} \and \author{Ben-Zhong, Dai\altaffilmark{2,3~\ddag}}
\email{\emaila}

\altaffiltext{1}{Department of Physics, Jinggangshan University, Ji'an, 343009, Jiangxi, China}
\altaffiltext{2}{Department of Astronomy, Yunnan University, Kunming, 650091, Yunnan, China}
\altaffiltext{3}{Key Laboratory of Astroparticle Physics of Yunnan Province, Yunnan University, Kunming, 650091, China}
\altaffiltext{\ddag}{\emaila}

\begin{abstract}
In the work, we explore the role of jet properties in the spectral evolution
for a sample of Fermi-LAT bright blazars composed primarily by flat spectrum radio quasars (FSRQs).
We introduce a near-equipartition log-parabola (NELP) model
to fit the quasi-simultaneous multi-waveband spectral energy distributions (SEDs). The Markov Chain Monte Carlo (MCMC) method
is employed to determine the best values of spectral parameters and its errors.
The correlations of synchrotron peak luminosity $L_{pk}^*$ and its peak frequency $\nu_{pk}^*$, SSC-dominant factor $\xi_s$ and $\nu_{pk}^*$, and their implications as to the spectral evolution were studied.
The statistical analysis and the comparison with theoretical approximation indicate that: (1)the spectral evolution in $L_{pk}^*-\nu_{pk}^*$ plane may not only ascribed to variation of the characteristic energy $\gamma_{br}^\prime$ of the employed electron energy distribution (EED), but that other quantities must be varying as well.(2) for the spectral evolution in $\xi_s-\nu_{pk}^*$ plane the magnetic field $B^\prime$
may play the dominant role, and $\gamma_{br}^\prime$ may be considered to be the subdominant role.
(3) Variation of the beaming factor $\delta_b$ may be mainly responsible for the absence of a possible correlation, since $B^\prime$ is strongly anti-correlated with the size of radiation zone $R_b^\prime$.
The relation is agreement with that derived from the analytical theory, and should be self-consistent with the underling acceleration mechanism of the employed EED. By assuming a conical jet with opening angle $\theta_{op}\sim1/\Gamma_b$,
we find that the $\gamma$-ray emission site being located at a
distance $\sim10^{17}-10^{19}$ cm.
\end{abstract}


\section{Introduction}
\label{sec:intro}

FSRQs are radio loud active galactic nuclei (AGNs), which typically exhibit substantial variability across the entire electromagnetic spectrum, from radio to $\gamma$-rays produced in a relativistic jet that is closely aligned with our line of sight \citep{Urry1995}.
Under the unification theory, they are classified along with BL Lac objects as blazars, and had been
revealed to be approximately equal with the number of BL Lac objects observed by the Fermi $\gamma$-ray Space Telescope.
In general, their spectral energy distribution (SED) is characterized by a typical double-peaked feature.
One component peaks in the infrared (IR) regime, and presents strong broad emission lines in optical bands,
while the second component extends from X-rays to $\gamma$-rays, and peaks in the high-energy range.
The lower-energy peak is interpreted as
synchrotron emission from a relativistic population of electrons, while the high-energy peak
is attributed to inverse-Compton up-scattering off of synchrotron photons emitted by the jet
itself, (SSC) or off of photons that are external to the jet (EC) \citep{GT2009,D2009, Hu2015}. A few plausible target photons for
the EC process are photons emitted by the accretion disk \citep{DS1997,DS2002}, radiation that is reprocessed and scattered off of
nearby clouds, and/or IR photons that scatter off of the dusty obscuring torus \citep{Sikora1994,SMM2000}. For HSP BL Lacs,
the homogeneous single zone SSC model can explain a given SED satisfactorily, however, one needs to invoke EC emission to explain the SEDs of FSRQs and LBL objects.

\cite{Fossati1998} combined several blazar surveys and noted an anti-correlation between $L_{pk}^*$ and $\nu_{pk}^*$.
It was claimed as evidence for a blazar sequence, which represents different blazar sub-classes, ranging from FSRQs through LBLs to HBLs.
The anti-correlation $L_{pk}^*-\nu_{pk}^*$ is related to the power injected into electrons and the break energy of the emitting electrons,
and then it is connected with mass of the central black hole and accretion ratio \citep[see][and references therein]{Finke2013}.
The author suggested that compton dominance is a more intrinsic indication of a blazar sequence,
since it is independent of the redshift of the source.
The author indicated that an anti-correlation between the compton dominance and $\nu_{pk}^*$ may exist for just BL Lac objects.
Since the peak Compton-scattered emission may be mainly contributed by the SSC process for BL Lac objects,
the Compton dominance is actually the SSC-dominant factor.

However, the SSC-dominant factor $\xi_s\equiv L_{pk}^{ssc}/L_{pk}^{syn}$ is much more difficult to determine for FSRQs through
the phenomenological approach, since the SSC component may be concealed by the EC component.
Moreover, it is more difficult to determine the quantities characterizing the synchrotron peak because
it typically occurs in the IR range, whereas most of the available data are almost entirely in the radio
and optical/UV bands. On the one hand, radio radiation emitted by the inner jet may be
opaque due to the synchrotron self-absorption process \citep{GT2009,GT2015,Ghisellini2010},
 and consequently the observed radio radiation may instead originate from
the extended jet \citep{Villata2006,Villata2007,Villata2009,Ogle2011}. On the other hand, the
optical/UV non-thermal continuum of some blazars may be contaminated by thermal
radiation emitted by the accretion disc and/or broad emission lines from the BLR \citep{Raiteri2007,Raiteri2011}.

Armed with the well-known radiative mechanisms, many authors utilized the simultaneous and/or quasi-simultaneous SEDs of multi-wavelength (MWL) to study the properties of blazar jets
in the framework of the one-zone leptonic model \cite[e.g.,][]{Ghisellini2010,Yan2012,Kang2014,Hu2017}.
In \cite{Kang2014}, the seed photons coming from the two different locations are respectively considered in the context of EC process. A $\chi^2$ test indicated that modeling with IR seed photons was systematically better than that with BLR photons.
Although a spectral break in the GeV spectrum revealed by the Fermi-LAT telescope was not taken into account,
they suggested that $\gamma$-ray-emitting regions are most likely found outside the BLR.
By modelling the MWL SEDs of 27 Blazars, \cite{Hu2017} proposed that the
SEDs modelling alone may not provide a significant constraint on the location of a high-energy emission region, where
both uncertainty of the $\gamma$-ray emission zones and the GeV spectrum break were considered.
Such a break could be arising from the combination of two components, namely, Compton-scattered disk and BLR photons \citep{Finke2010}, or from Klein-Nishina effects and a curving electron distribution \citep{Cerruti2013}.


Compared with previous works, our goal is to investigate the role of jet properties played in the spectral evolution for a sample of the Fermi-LAT bright blazars.
In our work, we modelled MWL SEDs with a physically motivated synchrotron/Compton model,
where the spectral parameters describing the MWL SEDs are directly related to the physical quantities describing
the physical properties of jets.
The spectral parameters and their corresponding errors are determined
via the Markov-Chain Monte-Carlo (MCMC) method, which is a powerful tool to systematically investigate high-dimensional
model parameter space \citep{Yan2013,Zhou2014,Peng2014,Yan2015}.
This paper is structured as follows. Section \ref{model} provides a brief description of the homogenous single-zone
leptonic model. Section \ref{fitting} shows our fitting procedures, and our results are reported in section \ref{resu}.
We discuss our results in Section \ref{diss}, 
and summarize our conclusions in Section \ref{summary}.
\section{The SED modelling}\label{model}
The blazar model used in this work is the generic leptonic jet model.
In the relativistic jet, a region (or blob) responsible for the emission of high energy $\gamma$-rays propagates at a relativistic speed $\beta_\Gamma=(1-1/\Gamma_b)^{1/2}$ outward along the jet, which is directed at an angle $\theta_{obs}$ with respect to the line of sight. Thus, the observed photons are beamed and Doppler-boosted towards the observer, where the Doppler boosting is determined by the Doppler factor $\delta_b=[\Gamma_b(1-\beta_\Gamma\cos\theta_{obs})]^{-1}$.
The high-energy emission region is modelled as a spherical magnetised plasma cloud of radius $R_b^\prime$
which consists of a population of isotropic relativistic electrons and a randomly oriented magnetic field $B^\prime$.
Electrons are emitted via synchrotron and IC mechanisms, which correspond to the low-energy and high-energy peaks of the observed SEDs, respectively.
The EED in the blob reference frame is assumed to be a Log-Parabola (LP) function
\begin{equation}
n_e^\prime(\gamma^\prime)= k_{br}^\prime\Big(\frac{\gamma^\prime}{\gamma_{br}^\prime}\Big)^{-s-r\log_{10}\frac{\gamma^\prime}{\gamma_{br}^\prime}},
\end{equation}
where $k_{br}^\prime$ is the normalization factor of the employed EED, the parameter $s$ is the spectral index at the characteristic energy $\gamma_{br}^\prime$ of the EED,  and $r$ measures the curvature of the function.
Here, $k_{br}^\prime$ is determined by the equilibrium relation $U_e^\prime=\xi_eU_B^\prime$, where $U_B^\prime$ and $U_e^\prime$ are
the magnetic-field and the electron-energy densities, respectively. The typical value of the ratio $\xi_e$ is ranging in $\simeq[120-180]$ for TeV HBLs,
while $\xi_e\simeq1$ is supported for the powerful FSRQs \citep[e.g.,][]{BottcherC2002,Bottcher2013,Kino2002,DS2002,SL2007,SL2008,Yan2016,Hu2017}.

\subsection{The Synchrotron Self-Compton process}

The observed synchrotron SED from isotropic electrons is given by
\begin{equation}
\nu F_{\nu}^{syn}=\frac{ V_b^\prime\delta_b^4}{4\pi d_L^2}\mathcal{J}_{sy}(\epsilon^\prime)\chi(\tau_{\epsilon^\prime}),
\end{equation}
where the synchrotron photon's dimensionless energy in the blob's frame $\epsilon^\prime$ is related to observed photon's frequency
by the relation $\epsilon^\prime m_ec^2/h=(1+z)\nu/\delta_b$, $V_b^\prime=4\pi R_b^\prime/3$ is the intrinsic volume of the blob and $d_L$ is the luminosity distance of source at redshift $z$.
Here $m_e$ is the rest mass of electron, $c$ is the speed of light, and $h$ is Planck's constant.

In the above equation, the synchrotron emissivity from isotropic electrons bathing in a randomly oriented magnetic field is given by
\begin{eqnarray}
\mathcal{J}_{sy}(\epsilon^\prime) &=& \frac{\sqrt{3}e^3B^\prime}{h}\epsilon^\prime\int_1^\infty d\gamma^\prime n_e^\prime(\gamma^\prime) R_s(\epsilon^\prime/\epsilon^\prime_c)
\end{eqnarray}
where $B^\prime$ is the magnetic field strength, $e$ is the fundamental charge, $\epsilon_c^\prime=\frac{3eB^\prime h}{4\pi m_e^2c^3}{\gamma^\prime}^2$ is the dimensionless characteristic energy of synchrotron radiation.
An accurate approximation of the function $R_s(x)$ is given by \cite{Finke2008}.
In the spherical approximation, the Synchrotron self-absorption (SSA) factor is $\chi(\tau)=1/2+\exp(-\tau)/\tau-[1-\exp(-\tau)]/\tau^2$, with the  opacity $\tau_{\epsilon^\prime}=2\kappa_{\epsilon^\prime} R_b$, where the dimensionless form of the SSA coefficient can be written as
\begin{equation}
\kappa_{\epsilon^\prime}=\frac{-\sqrt{3}B^\prime\lambda_c^3e^3}{8\pi hm_ec^3{\epsilon^\prime}^2}\int_1^\infty d\gamma^\prime R_s(\epsilon^\prime/\epsilon^\prime_c)\Big[{\gamma^\prime}^2\frac{\partial}{\partial\gamma^\prime}\Big(\frac{n_e^\prime(\gamma^\prime)}{{\gamma^\prime}^2}\Big)\Big],
\end{equation}
where $\lambda_c=h/m_ec=2.43\times10^{-10}~\rm cm$ is the electron Compton wavelength.

For the completely isotropic electrons and synchrotron photons, the observed SED of IC scattering can be written as
\begin{equation}
\nu F_{\nu}^{ssc}=\frac{{V_b^\prime}\delta_b^4}{4\pi d_L^2}\int_0^\infty{}d\epsilon^\prime[ u_{syn}^\prime(\epsilon^\prime)+u_{ssc}^\prime(\epsilon^\prime)]\mathcal{J}_{ic}(\epsilon^\prime,\epsilon_c^\prime),
\end{equation}
where the scattered photon energy in comoving frame is related to the observed photon frequency by the relation $\epsilon_c^\prime m_ec^2/h=(1+z)\nu/\delta_b$,
and the redistribution function is
\begin{equation}
\mathcal{J}_{ic}(\epsilon^\prime,\epsilon_s^\prime)=\frac{3}{4}c\sigma_T\int_1^\infty {}d\gamma^\prime{}n_e^\prime(\gamma^\prime)\mathcal{R}_{ic}(\epsilon^\prime,\gamma^\prime,\epsilon_s^\prime),
\end{equation}
and
\begin{eqnarray} \label{core}
\mathcal{R}_{ic}(\epsilon^\prime,\gamma^\prime,\epsilon_s^\prime)&=&\left(\frac{\epsilon_s^{\prime}}{\epsilon^\prime{\gamma^\prime}}\right)^2\Big[2x\ln{x}+x+1-2x^2\\
&+&\frac{(4\epsilon_s^\prime\gamma^\prime{}x)^2}{2(1+4\epsilon^\prime\gamma^\prime{}x)}(1-x)\Big]H(x;\frac{1}{4{\gamma^\prime}^2},1),\nonumber
\end{eqnarray}
with $x=\frac{\epsilon_s^\prime}{4\epsilon^\prime\gamma^\prime(\gamma^\prime-\epsilon_s^\prime)}$. Here, the energy densities of synchrotron and/or SSC radiation emitted by the same electrons are calculated through $u_{syn/ssc}^\prime(\epsilon^\prime)=\frac{4\pi d_L^2\nu F_\nu^{syn/ssc}}{4\pi R_b^{\prime2}cf_g\epsilon^\prime\delta_b^4}$ with $f_g=4/9$.
In the calculation, the SSC mechanism is calculated to an arbitrarily high scattering order, and synchrotron self-absorption is taken into consideration.

\subsection{The External Compton process}
In the work, BLR photons are characterized by 35 emission-line components and DT photons are assumed to have a Planckian distribution.
The relative intensities of the lines, expressed as the ratio of line fluxes compared to the strongest $Ly_\alpha$ emission line flux with a reference
value of 100, are presented in Figure 2 of \cite{Liu2006}. The sum of the line ratios is
equal to 555.77. Therefore, the energy density contributed by a certain emission line can
be normalized by $\rm U_{line}= N_{line}U_{BLR}/555.77$, where $N_{line}$ is the relative strength of a certain
emission line. For the DT, a fraction $f_{IR}=0.2$ of the accretion disk radiation is re-emitted as thermal IR radiation.
An effective temperature of the planckian distribution is expressed as $T_{DT}=[U_{DT}/a]^{1/4}$,
where the symbol $a$ denotes the radiation constant, and the IR photon field energy density of the DT is $U_{DT}=f_{IR}L_d/4\pi R_{IR}^2c ~\rm erg/cm^3$ with
 $R_{IR}=2.5\times10^{18}L_{d,45}^{1/2}$ cm indicated by near-IR interferometric measurements of the DT \citep{TG2008a,Landt2010,Malmrose2011}.

The $\nu F_\nu$ spectrum of Compton-scattered external isotropic radiation field can be written as
\begin{eqnarray}
\nu F_{\nu}^{iec}&=&\frac{{V_b^\prime}\delta_b^4}{4\pi d_L^2}\Big[\sum_{i=1}^{n=35}\frac{N_{line}U_{BLR}}{555.77}\mathcal{J}_{ic}(\epsilon^i_{line},\epsilon_c)\nonumber\\
&+&\int_0^\infty{}d\epsilon_*\frac{15U_{DT}\epsilon_*^3/(\pi\Theta)^4 }{\exp(\epsilon_*/\Theta)-1}\mathcal{J}_{ic}(\epsilon_*,\epsilon_c)\Big],
\end{eqnarray}
where $\epsilon_cm_ec^2=(1+z)h\nu$, and $\Theta=k_BT_{DT}/m_ec^2$ is the dimensionless temperature of the DT radiation field. Here, $\epsilon_c$ and $\epsilon_*$ refer to frame stationary with respect to the black hole,
the function $\mathcal{R}_{ic}(\epsilon_*,\gamma,\epsilon_c)$ is given by Eq.\ref{core}, and all primed quantities are replaced with the ones referring to the stationary frame.

\begin{table*}[t]
\centering
\tiny
\setlength{\tabcolsep}{2.0pt}
\def\arraystretch{1.5}
\caption{The subsample taken from the Fermi-LAT Bright AGN Sample \citep{abdo2010}.}
\begin{tabular}{lcccccccccc}
\hline
\hline
Name&$z$ &$\rm M_{BH}$ &$\rm L_d$ & $\rm \nu_{pk}L_{pk}$ & $\nu_{pk}$ & $\xi_s$ & r & $\rm t_{var}$  & $\rm U_{BLR}$ & $\chi^2(dof)$ \\
& &$\rm 10^8 M_\odot$ &$\rm 10^{46}\rm erg/s$ & $\rm erg/s$ & $\rm Hz$  & &  & $\rm sec$ & $\rm erg/cm^3$ &\\
(1) &(2) &(3) &(4) &(5) &(6)  &(7) &(8)  &(9)  &(10)  &(11)\\
\tableline
S4 0133\tablenotemark{\ddag} &0.859 &10 &1.1  &$1.03\pm0.03(10^{47})$  &$9.09\pm0.56(10^{13})$  &$0.71\pm0.20$  & $1.86\pm0.09$  &$3.94\pm2.47(10^4)$  &$3.52\pm1.23(10^{-4})$   &5.22(8)\\
PKS 0208\tablenotemark{\dag}&1.003 &7  &1.47 &  $4.90\pm0.45(10^{46})$ & $1.79\pm0.26(10^{13})$ & $0.41\pm0.07$ & $0.87\pm0.11$ & $6.51\pm0.29(10^5)$ & $2.76\pm0.77(10^{-4})$  & 4.40(14)\\
PKS 0227\tablenotemark{\dag}&2.115 &20 &3.0 &  $4.12\pm1.15(10^{47})$ & $2.02\pm0.36(10^{13})$ & $0.53\pm0.17$ & $3.28\pm0.67$ & $4.98\pm2.85(10^5)$ & $0.56\pm0.16(10^{-4}) $ & 1.23(10) \\
4C 28.07\tablenotemark{c}&1.213 &1  &0.15 &  $4.50\pm1.19(10^{46})$ & $1.98\pm0.43(10^{13})$ & $0.63\pm0.18$ & $1.14\pm0.24$ & $5.70\pm2.46(10^5)$  & $5.01\pm2.61(10^{-4})$ &2.15(3) \\
PKS 0347\tablenotemark{\dag}&2.944 &50 &7.5  &$2.53\pm0.36(10^{47})$  &$4.88\pm0.66(10^{13})$  &$0.79\pm0.21$  &$1.07\pm0.12$  &$20.39\pm8.92(10^5)$  &$14.43\pm9.67(10^{-6})$  &21.03(4)\\
PKS 0420\tablenotemark{a}&0.916 &1.1&0.43 &  $4.26\pm0.28(10^{46})$ & $0.79\pm0.10(10^{13})$ & $0.18\pm0.03$ & $0.47\pm0.03$ & $6.81\pm2.08(10^5)$ & $7.78\pm4.03(10^{-6})$ &3.24(14) \\
PKS 0454\tablenotemark{\ddag}&1.003 &30 &1.2 &  $2.75\pm0.44(10^{46})$ & $2.67\pm0.37(10^{13})$ & $0.38\pm0.10$ & $0.68\pm0.10$ & $13.13\pm4.62(10^5)$ & $4.86\pm2.24(10^{-4})$ &2.13(19) \\
PKS 0537\tablenotemark{\dag}&0.892 &20 &1.2  &  $7.06\pm0.52(10^{46})$ & $3.93\pm0.50(10^{13})$ & $0.92\pm0.26$ & $0.83\pm0.06$ & $13.34\pm9.37(10^5)$ & $4.07\pm1.84(10^{-4})$ & 3.26(20)\\
PKS 0851\tablenotemark{\dag}&0.306 &5  &0.015&  $3.11\pm0.10(10^{46})$ & $5.39\pm0.25(10^{13})$ & $0.64\pm0.29$ & $2.31\pm0.08$ & $1.00\pm1.00(10^5)$  & $0.47\pm0.15(10^{-4})$ &7.02(23) \\
4C 29.45\tablenotemark{\dag}&0.729 &10 &0.6  &$1.82\pm0.46(10^{46})$  &$1.42\pm0.36(10^{13})$  &$0.52\pm0.17$  &$1.05\pm0.25$  &$2.09\pm0.55(10^6)$  &$1.08\pm0.75(10^{-4})$  &0.89(9)\\
3C 279\tablenotemark{\dag}  &0.536 &8  &0.3  &$1.78\pm0.16(10^{46})$  &$1.44\pm0.22(10^{13})$  &$0.67\pm0.16$  &$0.57\pm0.06$  &$6.38\pm2.22(10^6)$  &$1.70\pm0.65(10^{-4})$   &5.10(35)\\
PKS 1454\tablenotemark{\ddag}&1.424 &20 &4.0  &$2.43\pm0.40(10^{47})$  &$2.67\pm0.24(10^{13})$  &$0.29\pm0.05$  &$1.96\pm0.27$  &$6.35\pm2.31(10^4)$  &$2.09\pm0.60(10^{-4})$ &9.07(9)\\
PKS 1502\tablenotemark{\ddag}&1.839 &30 &3.0 &$2.23\pm0.13(10^{47})$  &$3.58\pm0.20(10^{13})$  &$0.48\pm0.06$  &$0.88\pm0.06$  &$7.76\pm1.65(10^5)$  &$3.10\pm0.68(10^{-4})$  &9.19(15)\\
PKS 1510\tablenotemark{\dag}&0.36  &7  &0.42 &$3.55\pm0.29(10^{45})$  &$1.77\pm0.16(10^{13})$  &$1.46\pm0.23$  &$1.10\pm0.07$  &$54.29\pm21.89(10^5)$  &$64.84\pm19.79(10^{-4})$  &7.15(19)\\
B2 1520\tablenotemark{\dag} &1.487 &25 &0.56 &  $9.16\pm2.54(10^{46})$ & $1.86\pm0.36(10^{13})$ & $0.13\pm0.04$ & $2.86\pm0.58$ & $5.92\pm2.52(10^4)$ & $1.00\pm0.19(10^{-4})$ & 2.35(6) \\
4C 66.20\tablenotemark{\ddag}&0.657 &10 &0.5 &  $1.85\pm0.27(10^{46})$ & $7.10\pm1.02(10^{13})$ & $0.81\pm0.14$ & $0.73\pm0.11$ &$5.63\pm2.17(10^5)$ &$5.80\pm2.11(10^{-4})$ &1.83(6)\\
S3 2141\tablenotemark{b}&0.213 &1.4&0.1  &  $4.40\pm0.24(10^{45})$ & $3.84\pm0.57(10^{14})$& $1.62\pm0.28$ & $1.14\pm0.06$ & $3.95\pm1.08(10^4)$ & $5.95\pm2.29(10^{-4})$ & 0.72(5)\\
BLLAC\tablenotemark{\dag} & 0.069 & 5 &0.003  &$1.75\pm0.10(10^{45})$  &$1.54\pm0.24(10^{13})$  &$0.048\pm0.004$  &$0.68\pm0.04$  &$0.14\pm0.034(10^4)$  &$1.08\pm0.44(10^{-4})$  &6.84(32)\\
3C 454.3\tablenotemark{\dag}&0.859 &10 &3.0  &$2.91\pm0.16(10^{47})$  &$1.17\pm0.09(10^{13})$  &$0.52\pm0.07$  &$0.95\pm0.05$  &$17.81\pm6.31(10^5)$  &$0.69\pm0.11(10^{-4})$  &2.29(23)\\
PKS 2325\tablenotemark{\dag}&1.843 &10 &4.5  &$5.72\pm1.19(10^{46})$  &$0.27\pm0.04(10^{13})$  & $0.06\pm0.02$  &$0.59\pm0.05$  &$52.67\pm28.45(10^4)$  &$0.42\pm0.31(10^{-4})$  &1.20(13)\\
PMN 2345\tablenotemark{\dag}&0.621 &4  &0.36 &  $0.74\pm0.13(10^{46})$ & $0.83\pm0.17(10^{13})$ & $0.12\pm0.03$ & $1.67\pm0.24$ & $2.58\pm1.12(10^5)$ & $0.88\pm0.43(10^{-4})$ & 6.21(8)\\
\hline
\hline
\end{tabular}
\tablecomments{Columns 1 and 2 are the source name and the redshift, respectively.
The mass of the black hole and the accretion disk luminosity are respectively shown in Columns 3 and 4.
Columns 5-10 are the input parameters (See the text for details.). The last column is the reduced $\chi^2$, and we present the degree of freedom in parenthesis.}
\tablenotetext{\ddag}{The mass of the black hole are derived from the SED modelling}
\tablenotetext{\dag}{The mass of the black hole are selected from \cite{Ghisellini2010}}
\tablenotetext{a}{The accretion disk luminosity is selected from \cite{Gu2001}}
\tablenotetext{b}{The accretion disk luminosity is selected from \cite{Paltani2005}}
\tablenotetext{c}{The accretion disk luminosity is selected from \cite{Fan2009}}
\label{input}
\end{table*}
\section{The fitting procedure}\label{fitting}
In the fitting procedure, the input parameters are comprised of the synchrotron luminosity ($\nu_{pk}L_{pk}$), its peak frequency ($\nu_{pk}$), a SSC-dominant factor ($\xi_s$),
the curvature parameter ($r$) of the EED, the minimum timescale of variation ($t_{var}$), and the BLR radiation energy densities ($U_{BLR}$).
Since $k_{br}^\prime$ is coupled with $B^\prime$,
the interesting physical parameters considered in the model are composed by the magnetic field ($B^\prime$), the Doppler factor ($\delta_b$),
the characteristic energy ($\gamma_{br}^\prime$), and the size of a radiation zone ($R_b^\prime$).
The relevant physical parameters are related to
the input parameters through the analytic relationships Eqs.(\ref{analy1},\ref{analy2},\ref{analy3},\ref{analy4}) presented in Appendix \ref{derivations}.
Compared with previous works, we relax the index $s$ as a free parameter and take into account the effects of cosmological expansion.
Throughout this work, a standard cold dark matter ($\rm \Lambda CDM$) cosmology is adopted, with a
Hubble constant of $H_0=71 \rm km/s/Mpc$, $\Omega_m=0.27$ and $\Omega_\Lambda=0.73$ as derived from Wilkinson
Microwave Anisotropy Probe results \citep{Spergel2007}. We use the notation $\rm Q=Q_x10^x$ in cgs units.

In particular,we set $s=2$, because the LP EED can be determined by the frequency peak position $\nu_{pk}$ and the corresponding peak luminosity $\nu_{pk}L_{pk}$ of the synchrotron SEDs.
Since our interesting is focused on the powerful blazars, we forced that $\xi_e=1$.
To match the day-scale variability seen in most Fermi-detected FSRQs, we let $t_{var}=8.64\times10^{4}$ as an initial value.
Subsequently, we can obtain the plausible values of $(\nu_{pk}, \nu_{pk}L_{pk},r)$ and $\xi_s$, from the low-energy part and $X-$ray band of the observed SEDs, respectively.
The plausible values of $U_{BLR}$ can be derived by fitting the high-energy part of the observed SEDs.
Finally, we employed the MCMC method instead of a simple $\chi^2$ minimization strategy to obtained the best fit and the corresponding errors of the input parameters.
The MCMC code we used \citep[e.g.,][]{Liu2012,Yuan2011} was adapted from COSMOMC. Thus we refer the reader
to \cite{Lewis2002} for a detailed explanation of the code about sampling options, convergence criteria, and statistical quantities.


\section{Results}\label{resu}
\begin{table}[t]
\centering
\tiny
\setlength{\tabcolsep}{2.0pt}
\def\arraystretch{1.5}
\caption{Summary of results for the derived physical quantities.}\label{output}
\begin{tabular}{lcccc}
\hline
\hline
Name &$B^\prime$ &$\delta_b$ &$\gamma_{br}^\prime$ &$R_{b}^\prime$ \\ 
\hline
S4 0133 & $1.35\pm0.62$ & $22.12\pm3.23$ & $624.79\pm152.04$ &$1.41\pm0.79\cdot10^{16}$\\ 
PKS 0208& $0.38\pm0.06$ & $13.98\pm1.12$ & $334.77\pm66.50$ &$1.36\pm0.12\cdot10^{17}$\\ 
PKS 0227& $0.37\pm0.17$ & $21.82\pm3.67$ & $956.40\pm230.60$ &$1.05\pm0.55\cdot10^{17}$\\ 
4C 28.07& $0.51\pm0.19$ & $12.30\pm1.81$ & $470.43\pm140.74$ &$9.51\pm3.81\cdot10^{16}$\\ 
PKS 0347& $0.21\pm0.08$ & $16.90\pm2.24$ & $1214.40\pm270.82$&$2.62\pm1.05\cdot10^{17}$\\ 
PKS 1420& $0.37\pm0.09$ & $16.26\pm1.38$& $66.95\pm14.25$  &$1.73\pm0.48\cdot10^{17}$ \\ 
PKS 0454& $0.23\pm0.08$ & $12.27\pm1.58$ & $388.71\pm123.90$ &$2.41\pm0.8\cdot10^{17}$\\ 
PKS 0537& $0.34\pm0.17$ & $10.50\pm1.60$ & $552.71\pm163.14$ &$2.22\pm1.39\cdot10^{17}$\\ 
PKS 0851& $0.81\pm0.58$ & $14.02\pm3.28$ & $738.17\pm285.17$ &$3.19\pm2.88\cdot10^{16}$\\ 
4C 29.45& $0.23\pm0.08$ & $8.61\pm1.36$  & $573.28\pm187.5$ &$3.11\pm0.87\cdot10^{17}$\\ 
3C 279  & $0.19\pm0.06$ & $6.13\pm0.72$  & $288.02\pm79.73$ &$7.64\pm2.47\cdot10^{17}$\\ 
PKS 1454& $0.77\pm0.21$ & $33.20\pm2.94$ & $431.07\pm73.3$ &$2.61\pm0.85\cdot10^{16}$\\ 
PKS 1502& $0.29\pm0.05$ & $20.39\pm1.27$ & $549.66\pm72.25$ &$1.67\pm0.32\cdot10^{17}$\\ 
PKS 1510& $0.27\pm0.08$ & $3.47\pm0.30$  & $864.47\pm156.46$ &$4.16\pm1.50\cdot10^{17}$\\ 
B2 1520 & $0.47\pm0.18$ & $38.25\pm5.85$ & $524.23\pm120.0$ &$2.73\pm1.07\cdot10^{16}$\\ 
4C 66.20& $0.47\pm0.14$ & $9.90\pm0.96$  & $510.70\pm163.18$ &$1.01\pm0.35\cdot10^{17}$\\ 
S3  2141& $1.92\pm0.44$ & $9.15\pm0.78$  & $916.33\pm141.54$ &$8.92\pm2.25\cdot10^{15}$\\ 
BLLAC  & $3.89\pm0.70$ & $33.98\pm1.79$& $31.57\pm5.59$  &$1.33\pm0.29\cdot10^{15}$\\ 
3C 454.3& $0.24\pm0.06$ & $14.67\pm1.10$ & $363.25\pm55.71$ &$4.21\pm1.33\cdot10^{17}$\\ 
PKS 2325& $0.26\pm0.11$ & $27.63\pm4.75$ & $71.75\pm21.6$  &$1.53\pm0.79\cdot10^{17}$\\ 
PMN 2345& $0.28\pm0.10$ & $16.71\pm2.14$ & $415.85\pm100.3$ &$7.97\pm3.15\cdot10^{16}$\\ 
\hline
\hline
\end{tabular}
\end{table}

In this section, we apply our model-fitting procedure to a set of $\gamma$-ray blazars detected by Fermi-LAT.
The SEDs were taken from \cite{abdo2010}, where MWL quasi-simultaneous SED of 48 bright blazars in the Fermi-LAT Bright AGN Sample
were reported.
Out of their list, we selected a subset of blazars which is mainly comprised of the powerful FSRQs.
For the sample blazars, the accretion disk luminosity ($L_d$) and the mass of the black hole $M_8$ were either collected from the literature,
or they were derived by attempting to model the optical/UV data with a standard Shakura-Sunyaev disk \citep{SS1973}. The archival data at low state, taken from the NASA Extragalactic Data base, were available.
For some sources, a relative systematic uncertainty of 5\% was assumed due to the unavailable errors associated with the optical, UV, and X-ray photons.
The extragalactic background light (EBL) model of \cite{Finke2010} was used to correct for the absorption affect,
and was modified by normalization factor of 0.86 \citep{Ackermann2012}.
The observed SEDs of our sources, together with the best-fitting model SEDs, are shown in Figures \ref{figsed1}-\ref{figsed3}.
The mean values and standard deviation of marginalized probability posterior of the input parameters 
are summarized in Table \ref{input}, while in Table \ref{output} we list the values of the derived parameters and their $1\sigma$ errors
derived from the standard error propagation formula.

\begin{figure}[t]
  \centering
  \includegraphics[width=8cm]{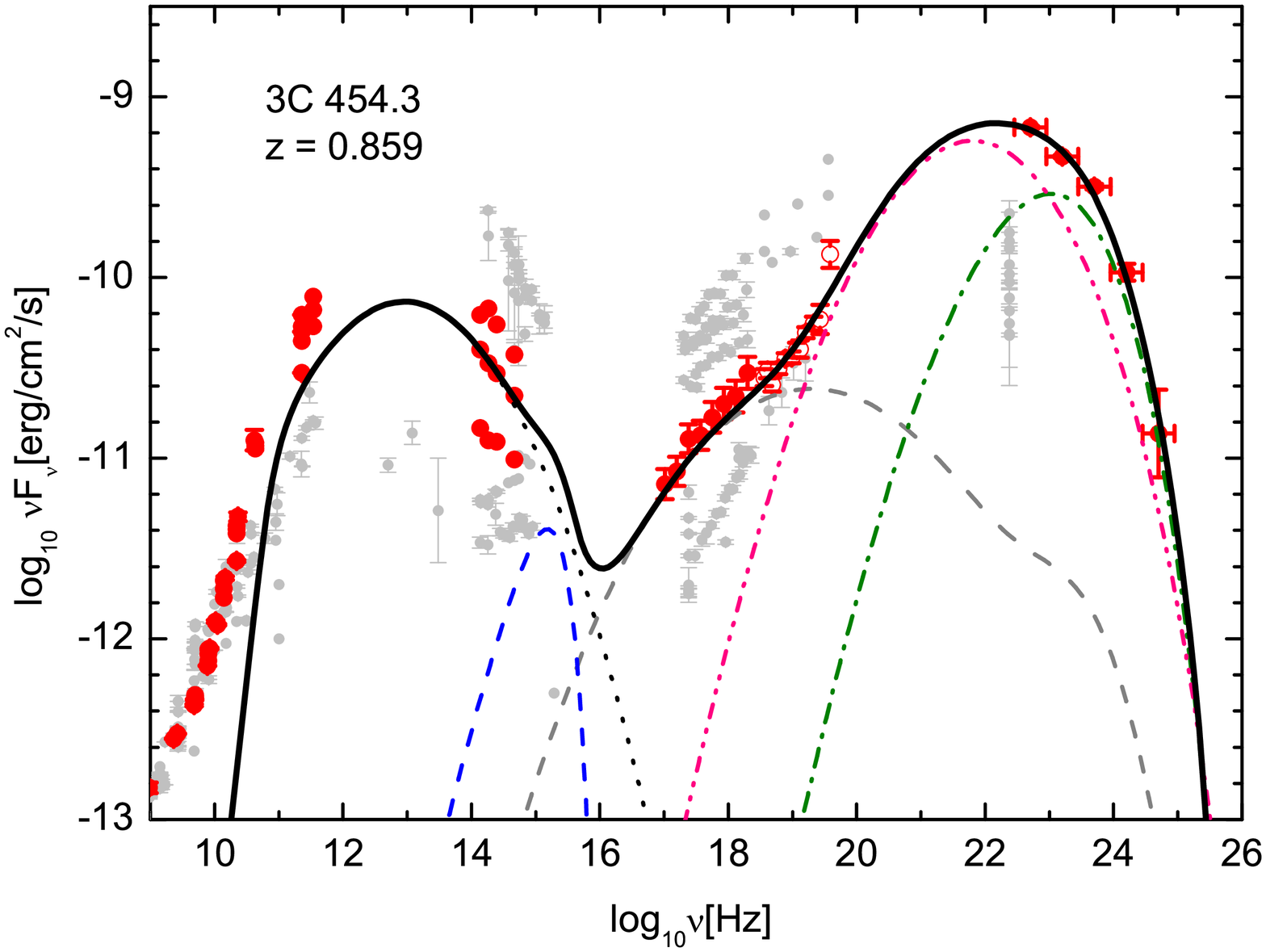}\\
  \includegraphics[width=8cm]{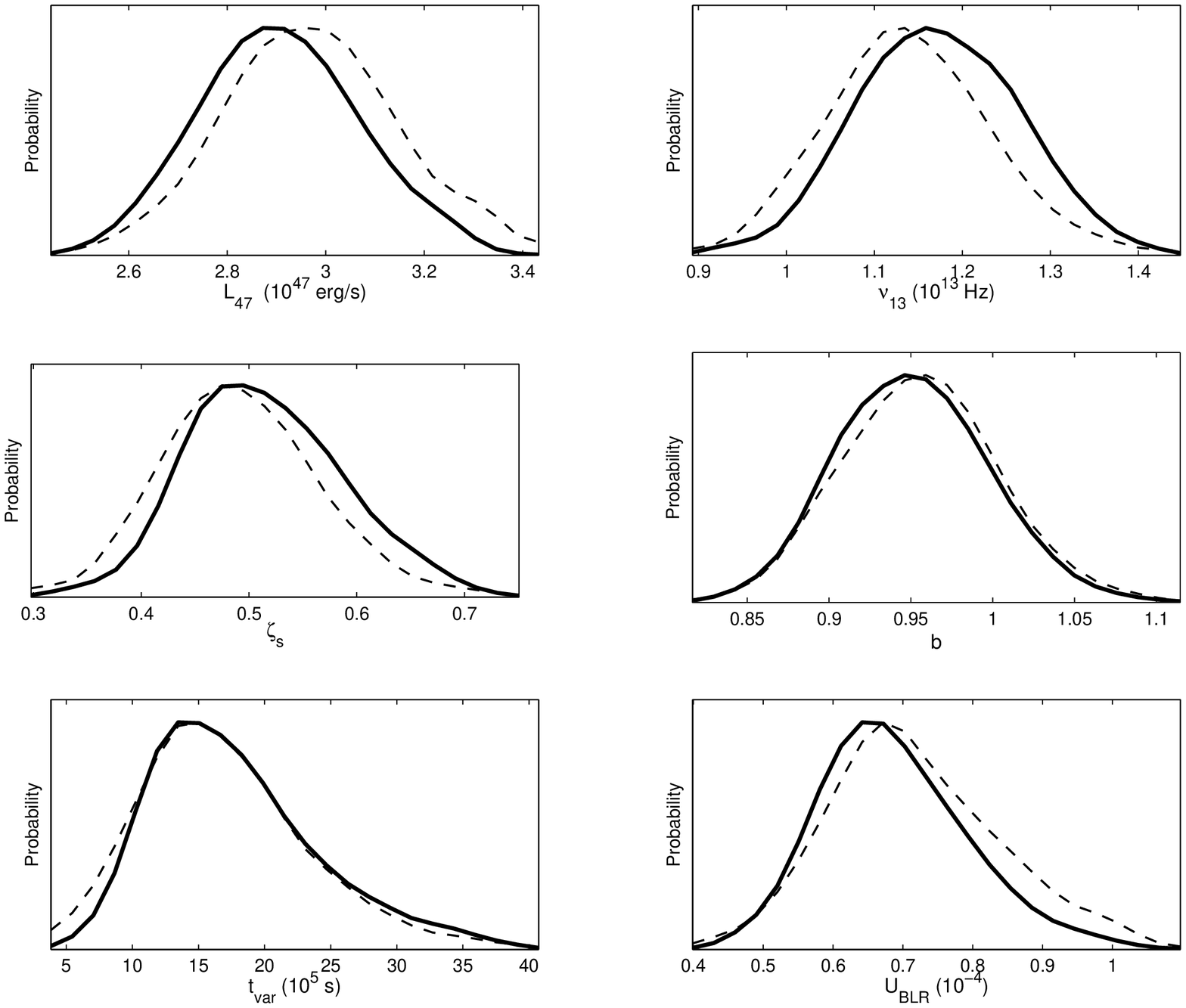}
  \caption{An example of the SED modelling and marginalized probability posterior of the parameters in SED modeling.
  Upper: Comparisons of the modelling SED with the quasi-simulation MWL SED data;
  Bottom: 1-D marginalized probability posterior (solid line) and relative mean likelihood (dash line) of the parameters.
 }\label{distr}
\end{figure}

Our results show that the model gave a satisfactory description of the SEDs and the spectral break in the GeV band for the majority of the sources.
For example, Figure ~\ref{distr} shows the comparisons of the modelling SED with the observed data, and the
probability distribution of each parameter for 3C 454.3, where $\nu_{pk}L_{pk}=2.91\pm0.16(10^{47})$, $\nu_{pk}=1.17\pm0.09(10^{13})$,
$\xi_s=0.52\pm0.07$, $r=0.95\pm0.05$, $t_{var}=17.81\pm6.31(10^5)$, $U_{BLR}=0.69\pm0.11(10^{-4})$.

\begin{table}[t]
\centering
\tiny
\setlength{\tabcolsep}{4.0pt}
\def\arraystretch{1.5}
\caption{Statistics of Correlations involving $L_{pk}^*$ and $\xi_s$.}\label{rela}
\begin{tabular}{lcc|cc}
\hline
\hline
Relations &r & p-value &r\tablenotemark{a}  &p-value\tablenotemark{a} \\
\hline
$\log\nu_{pk}^*-\log L_{pk}^*$ &0.21 &0.19  &0.48  &0.04 \\
$\log\nu_{pk}^*-\log\xi_s$     &0.66  &0.24  &0.65  &0.29 \\
\hline
$\log R_b^\prime-\log B^\prime$&-0.95  &$7.92\times10^{-11}$  &-0.92  &$3.63\times10^{-7}$ \\
$\log R_{diss}-\log B^\prime$  &-0.96  &$2.0\times10^{-10}$  &-0.94  &$4.87\times10^{-7}$ \\
\hline
$\log L_{pk}^*-\log\gamma_{br}^\prime$ &0.30  &0.04  &0.008  &0.97 \\
$\log L_{pk}^*-\log\delta_b$ & 0.43 &0.62  &0.69  & $3.95\times10^{-5}$\\
$\log L_{pk}^*-\log B^\prime$ & -0.37 & $1.1\times10^{-3}$ &-0.1  &0.16 \\
$\log L_{pk}^*-\log R_b^\prime$ &0.31  &$2.3\times10^{-3}$  &0.009  &0.37 \\
\hline
$\log\gamma_{br}^\prime-\log\xi_s$ &0.79  & $1.84\times10^{-5}$ &0.7  &$3.8\times10^{-3}$ \\
$\log\delta_b-\log\xi_s$ &-0.70  & $3.18\times10^{-6}$ & -0.66 &$1.88\times10^{-4}$ \\
$\log B^\prime-\log\xi_s$ &-0.15  &$6.3\times10^{-3}$  &0.26  &0.052 \\
$\log R_b^\prime-\log\xi_s$ &0.35  &$1.8\times10^{-4}$  &0  &0 \\
\hline
\end{tabular}
\tablenotetext{a}{The values of $r$ and $p$ are calculated by excluding BLLAC.}
\end{table}

\begin{figure}[t]
  \centering
  \includegraphics[width=8cm]{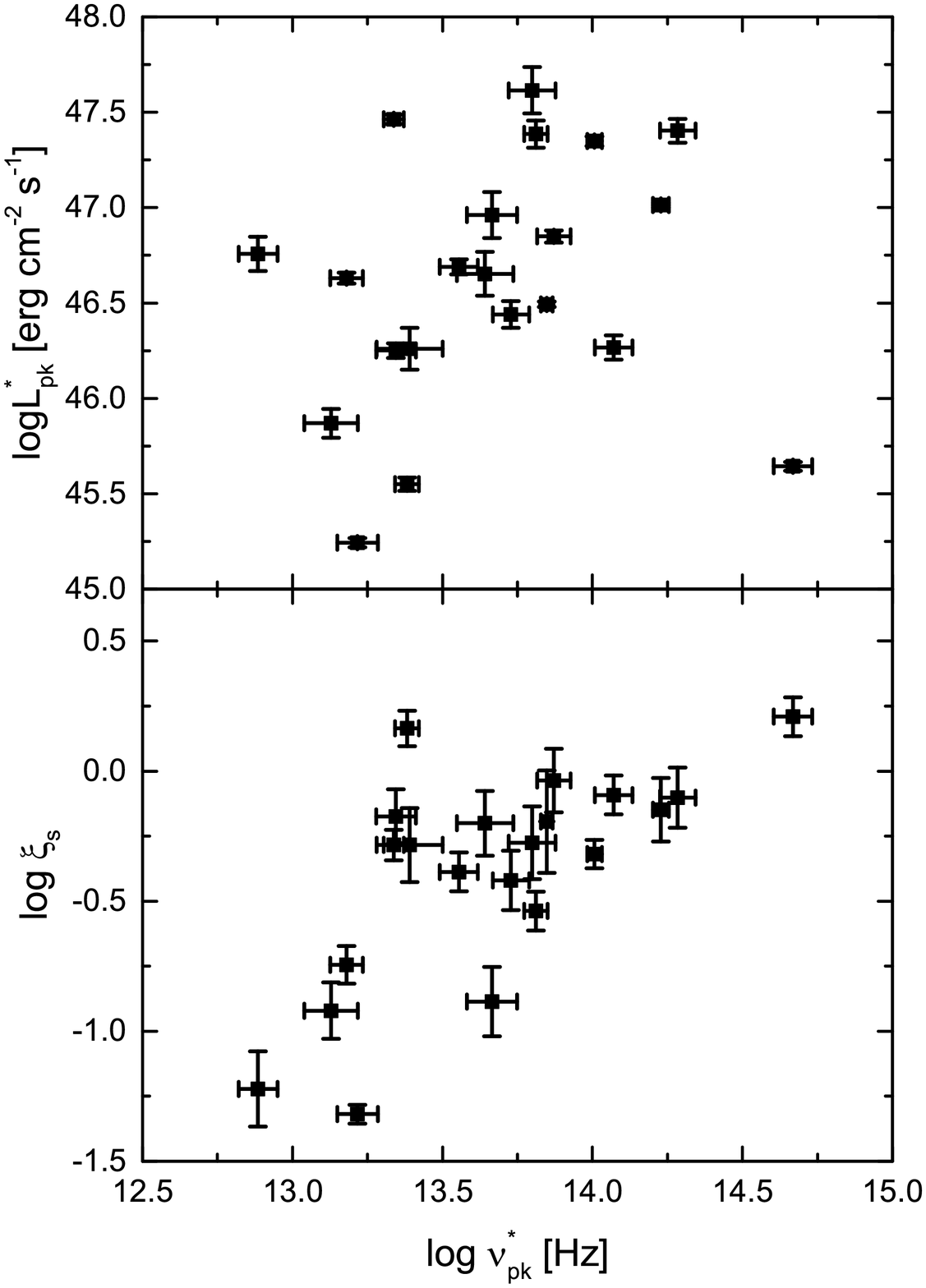}\\
  \caption{Top panel: the synchrotron peak luminosity $L_{pk}^*$ as a function of the corresponding frequency peak position $\nu_{pk}^*$.
  Bottom panel: the SSC-dominant factor as a function of $\nu_{pk}^*$.}\label{figure2}
\end{figure}

By using the values of the spectral parameters, we therefore make a statistical analysis
on the correlations or trends in the logarithms of spectral parameters ($L_{pk}^*,\xi_s,\nu_{pk}^*$),
where $L_{pk}^*\equiv\nu_{pk}L_{pk}$, and the rest-frame peak synchrotron frequency is given by $\nu_{pk}^*=(1+z)\nu_{pk}$ in order to perform an effective comparison.
The resulting scatter plot of $\log\nu_{pk}^*$ versus $\log L_{pk}^*$ is shown in the upper panel of Figure ~\ref{figure2}.
and the scatter plot of $\log\xi_s$ versus $\log\nu_{pk}^*$ is presented in the bottom plane of Figure ~\ref{figure2}.
To determine the strength of the correlations, we have computed the Pearson correlation coefficient, $r$, and the probability of no correlation, $p-value$. The results can be found in Table \ref{rela}. 
In the next section, we will discuss the matter in detail.

\section{Discussion and Conclusion}\label{diss}
\begin{figure}[t]
  \centering
  \includegraphics[width=8cm]{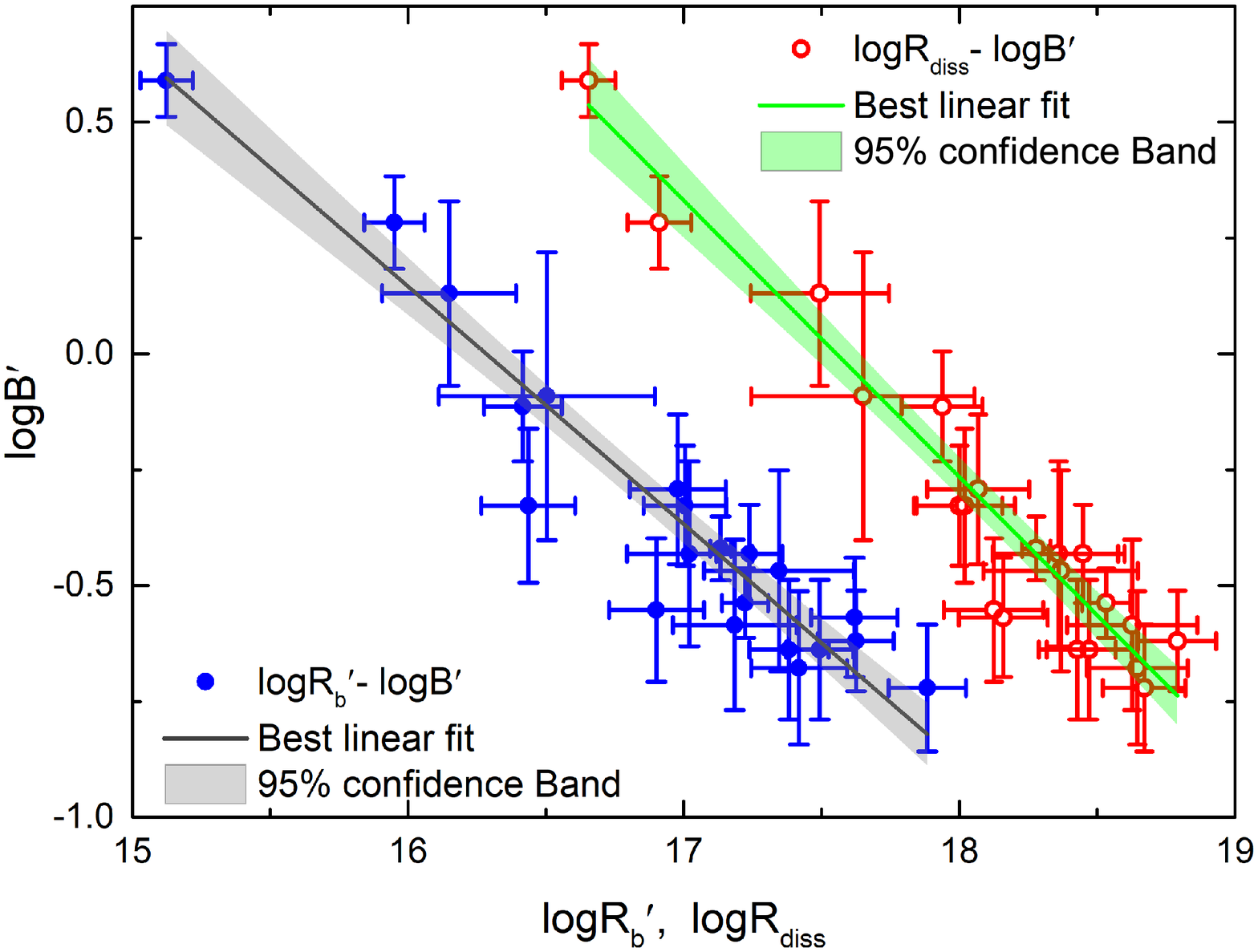}
  \caption{The magnetic-field strength $B^\prime$ as a function of the size of a radiation zone and of the location of a radiation zone.}\label{figure3}
\end{figure}

In the work, the collected SEDs represent an average emission state, and then the derived parameters only represent the average properties of the emission region.
We obtain values of $\delta_b$ mainly ranging between $10-40$, and values of $\gamma_{br}^\prime$ mainly ranging between $300-1000$.
$B^\prime$ and $R_b^\prime$ are values of the order of 0.2 - 1.0 G, and $10^{16}-10^{17}$ cm, respectively.
Further, a closer analysis of a possible correlation or trend among the quantities indicated that a strong anti-correlation between $B^\prime$ and $R_b^\prime$ are presented, and is shown in Figure ~\ref{figure3}. The Pearson test presents they are tightly correlated with a chance probability of $p=7.92\times10^{-11}$.
The best-fitting linear model gives $B^\prime\propto R_b^{\prime-0.51\pm0.04}$.
The statistically significant correlation is in good agreement with the theoretical prediction based on the jet model.
By eliminating $t_4$ between Eq.(\ref{analy1}) and Eq.(\ref{analy4}), we can obtain
\begin{equation}\label{BRR}
B^\prime\propto\left(\frac{f_1f_3}{f_2}\right)^{4/7}\frac{\left(f_0L_{pk}^*\right)^{1/14}}{{\nu_{pk}^*}^{2/7}}\xi_s^{1/2}R_b^{\prime-5/7}.
\end{equation}
This shows that the magnetic field falls more slowly
with the transverse size of jet than a naive scaling. If the magnetic field is mainly uniform and unidirectional, then the magnetic flux presumably would be a conserved quantity.
Then the magnetic-field strength $B^\prime\propto R_b^{\prime-2}$, falls off very quickly with distance from the power house.
For a completely tangled magnetic field, both observationally and in most theoretical models, $B^\prime\propto R_b^{-1}$ \citep{Begelman1984}.
Since the magnetic-field strength can be amplified by astrophysical shocks
beyond simple shock compression through a turbulent dynamo action \citep{Guo2012,Fraschetti2013}.
Thus, the slowly varying magnetic field may support the concept of magnetic-field amplification in relativistic jets of the powerful blazars, and should be self-consistent with stochastic acceleration scenarios that can naturally produce a log-parabolic EED.

Assuming a conical jet with opeing angle $\theta_{op}\sim1/\Gamma_b$,
the location of the $\gamma$-ray emitting region can be evaluated through the relation $R_{diss}\simeq ct_{var}\delta_b^2/(1+z)$.
Then, we obtain values of $R_{diss}$ mainly ranging between $10^{17}-10^{19}$ cm, and the scatter plot of $\log B^\prime-\log R_{diss}$ is presented in Figure.\ref{figure3}.
The best-fitting linear model gives $\log B^\prime=(10.47\pm0.89)+(-0.6\pm0.05)\log R_{diss}$ with a chance probability of $p=2.0\times10^{-10}$.
The result shows that the $\gamma$-ray emission site being located at a distance $R_{BLR}\lesssim R_{diss}<R_{IR}$, where $R_{IR}=2.5\times10^{18}L_{d,45}^{1/2}$ and $R_{BLR}=10^{17}L_{d,45}^{1/2}$ cm.
In addition, the energy densities of the external BLR photons field cluster around the order of $10^{-4} \rm erg/cm^3$.
The values we derive in our modeling are two orders of magnitude lower than the estimated values $0.03 ~erg/cm^3$ inside a BLR\cite[see e.g.][]{Ghisellini2010,Ghisellini2014}.
Thus, the energy densities of the external BLR photons field are consistent with the locations of the $\gamma-$ray emission region determined by assuming a conical jet.

For the aim of the present work, in the rest of this section we can directly get insight into the main driver of the spectral evolution
by combining the physical quantities characterised the properties of the jets and the spectral parameters characterised the MWL SEDs.

\subsection{The relations in the $\log\nu_{pk}^*-\log L_{pk}^*$ plane}

\emph{In the diagram of $\nu^*-\nu^* L^*_\nu$, the peak synchrotron value can be expressed by the following relationship:}
\begin{equation}\label{delsyn}
L_{pk}^*=\frac{2}{3}c\sigma_TU_B^\prime V_b^\prime k_{pk}^\prime\delta^4=\frac{c\sigma_T\xi_ef_2}{96\pi^2m_ec^2I_1}V_b^\prime{\gamma_{br}^\prime}{B^\prime}^4\delta_b^4,
\end{equation}
where $k_{pk}^\prime=k_{br}^\prime\gamma_{br}^{\prime3}f_2$ represents the peak of $\gamma^{\prime3}n_e^\prime(\gamma^\prime)$,
and the corresponding peak Lorentz factor is $\gamma_{pk}^\prime=\gamma_{br}^\prime f_3$.
In the view of analytical theory, the spectral parameter $L^*_{pk}$ can be directly related
to variations in the physical parameters considered in this study. 
In the form, the dependence of $L_{pk}^*$ on $\nu_{pk}^*$ can be expressed as: $L_{pk}^*\propto{\nu_{pk}^*}^\beta$, in which the peak synchrotron
frequency can be approximated as $\nu_{pk}^*=\frac{3}{2}\gamma_{pk}^{\prime2}\epsilon_B^\prime\delta_bm_ec^2/h$.
Therefore, the power-law index $\beta$ can reflect changes of a mainly physical property of the blazar jets.
If changes in the characteristic energy $\gamma_{br}^\prime$ dominate, while the other physical quantities remain almost constant, we obtain
$\ln L_{pk}^*\propto\ln[\xi_eV_b^\prime(B^\prime\delta_b)^{7/2}]+0.5\ln\nu_{pk}^*$.
If variations in either $\delta_b$ or $B^\prime$ dominate, we also expect
$\ln L_{pk}^*\propto\ln[\xi_eV_b^\prime/\gamma_{br}^{\prime7}]+4\ln\nu_{pk}^*$.
Inversely, the dominant variation in $B^\prime$ or $\delta_b$ can not be
distinguished in the $\log\nu_{pk}^*-\log L_{pk}^*$ plane if the equipartition condition holds.
Formally, $\beta=\infty$ applies for any variations in the ratio $\xi_e$ between the non-thermal electron and magnetic-field energy densities,
or in the source volume, whereas it relates to changes only in the source volume if $\xi_e$ is constant.

\begin{figure}[t]
  \centering
  \includegraphics[width=8cm]{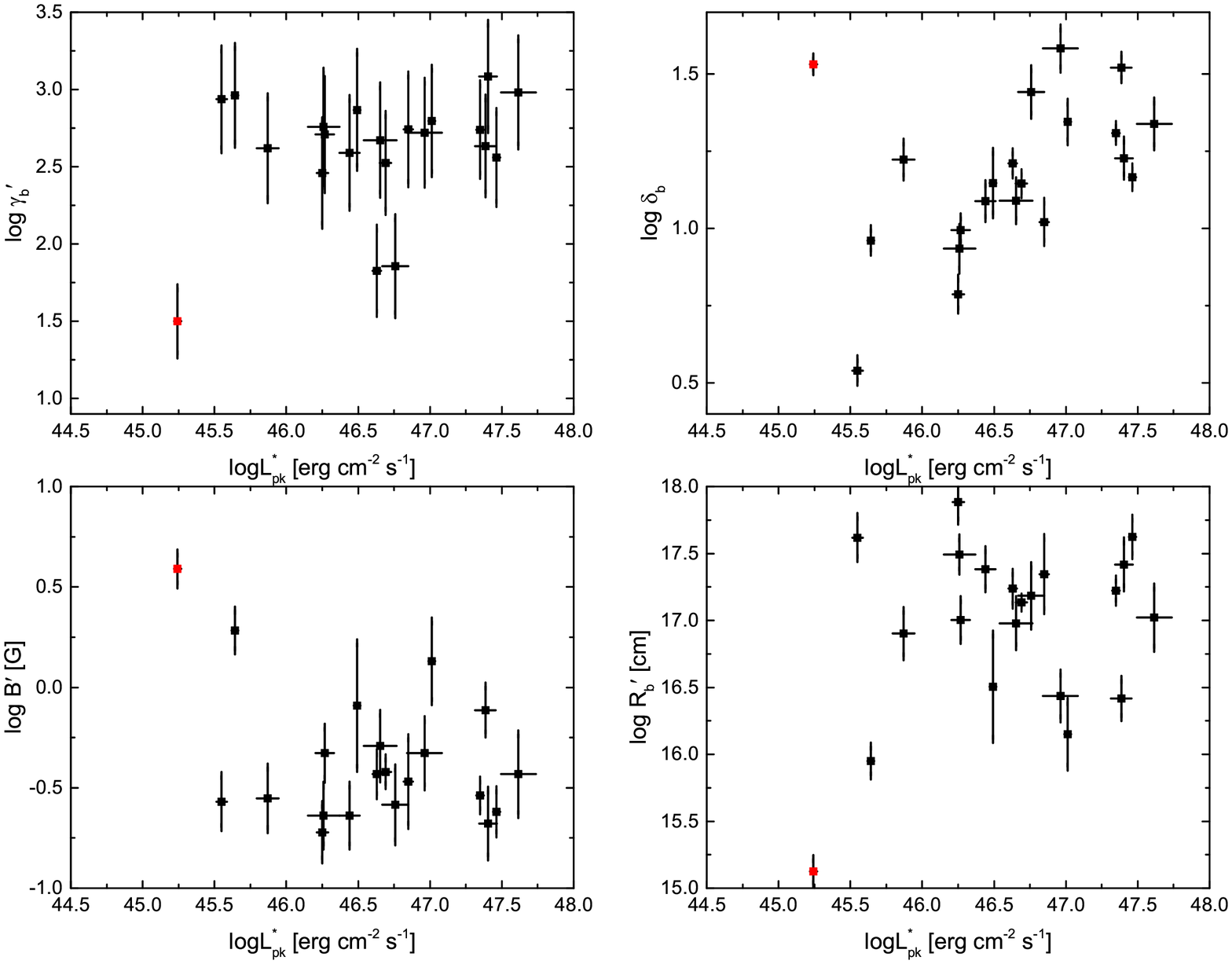}\\
  \caption{$\log\gamma_{br}^\prime-\log L_{pk}^*$, $\log\delta_b-\log L_{pk}^*$,
$\log B^\prime-\log L_{pk}^*$ and $\log R_b^\prime-\log L_{pk}^*$ plots for the sample. The red point represents the source BLLAC.
}\label{figure5}
\end{figure}

In the log-log representation, 
the resulting scatter plots of $\gamma_{br}^\prime$ versus $L_{pk}^*$, $\delta_b$ versus $L_{pk}^*$,
$B^\prime$ versus $L_{pk}^*$ and $R_b^\prime$ versus $L_{pk}^*$ are shown in Figure ~\ref{figure5}, and the Pearson correlation coefficient and $p-$values are reported in Table \ref{rela}.
The results show no correlation in the plot of $\log\delta_b-\log L_{pk}^*$, but three very weak correlations
may present in the $\log\gamma_{br}^\prime-\log L_{pk}^*$, $\log R_b^\prime-\log L_{pk}^*$ and $\log B^\prime-\log L_{pk}^*$ planes.
Through a jackknife statistical test, the three possible correlations are mainly contributed by BLLAC (marked as a red point).
Excluding this object, $L_{pk}^*$ and $\delta_b$ are correlated with a p-value $p=3.9\times10^{-5}$,
whereas no correlation exists between $L_{pk}^*$ and other physical quantities.
The absence of those correlations imply that the spectral evolution may be not only due to variations
of a single physical quantity but that other quantities must be varying as well.
The presence of the significant correlation implies $\delta_b$ is unlikely to be a dominant driver of the spectral evolution.
Since both $\nu_{pk}^*$ and $L_{pk}^*$ depend on $\delta_b$, the expected trend in the $\log\nu_{pk}^*-\log L_{pk}^*$ plane can be destroyed. 

On the other hand, if changes in $\gamma_{br}^\prime$ dominate,
$L_{pk}^*$ is expected to scale as $L_{pk}^*\propto\nu_{pk}^{*1/2}$ based on the above theoretical framework;
 meanwhile it is strongly dependent on $B^\prime$, $\delta_b$ and $R_b^\prime$ through
the relation $L_{pk}^*\propto R_b^{\prime3}\left(B^\prime\delta_{b}\right)^{7/2}$.
By substituting Eq.(\ref{BRR}), we can obtain $L_{pk}^*\propto R_b^{\prime1/2}{\delta_{b}}^{7/2}$, or $L_{pk}^*\propto B^{\prime-7/10}{\delta_{b}}^{7/2}$, if $\nu_{pk}^*$ occurs a small variation.
In fact, the values of $\nu_{pk}^*$ are mainly restricted within a narrow limit $[10^{13},10^{14}]$ Hz for the sample.
Furthermore, if $R_b^\prime$ or $B^\prime$ remains constant, $L_{pk}^*$ turns out to be proportional to $\delta_b^{3.5}$.
For the sample, our fit gives $\delta_b\propto L_{pk}^{*0.3}$ when excluding BLLAC.
The result is agree with the naive case, and implies that $\gamma_{br}^\prime$ may play a dominant role.

Motivated by the all above results, we propose that $\gamma_{br}^\prime$ may be the main driver of the spectral evolution for the sample.
The expected relation in the $\log\nu_{pk}^*-\log L_{pk}^*$ plane may be accompanied by a rescaling of a secondary role.
For the given sample, we does not make clear whether the secondary role of the spectral evolution is $B^\prime$ or $R_b^\prime$.
Since $B^\prime$ is tightly anti-correlated with $R_b^\prime$, $\delta_b$ may be mainly responsible for the absence of a possible relation.

\subsection{The relations in the $\log\nu_{pk}^*-\log\xi_s$ planes}
In the Thomson region, the SSC-dominant factor can approximately represent the ratio $\xi_s$
between the synchrotron radiation and magnetic-field energy densities in the co-moving frame, and is given by
\begin{equation}\label{mag1}
\xi_s=\frac{\sigma_TR_b^\prime}{18\pi m_ec^2}\frac{f_2\xi_e}{f_1f_g}\gamma_{br}^\prime B^{\prime2}.
\end{equation}
Obviously, it is independent with the beaming factor $\delta_b$.
If $\gamma_{br}^\prime$ is the change of a mainly physical quantity, the dependence of $\xi_s$ on $\nu_{pk}^*$ can
be represented as $\ln\xi_s\propto\ln[\xi_eR_b^\prime B^{\prime3/2}/\delta_b^{1/2}]+0.5\ln\nu_{pk}^*$.
It shows that the power-law dependence of $\xi_s$ on $\nu_{pk}^*$ is the same as that of $L_{pk}^*$ on $\nu_{pk}^*$
when changes of $\gamma_{br}^\prime$ dominate.
And we obtain $\ln\xi_s\propto\ln[\xi_eR_b^\prime/\delta_b^2\gamma_{br}^{\prime3}]+2\ln\nu_{pk}^*$,
when the change of a physical quantity is dominated by variations of $B^\prime$.
Formally, if changes in $\xi_e$ or the source volume dominate, $\beta=\infty$ can be expected.

\begin{figure}[t]
  \centering
  \includegraphics[width=8cm]{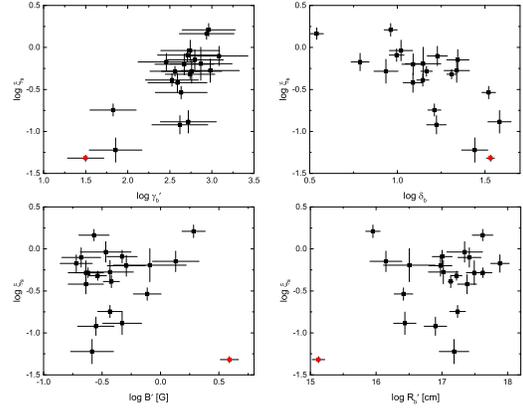}
  \caption{$\log\xi_s-\log\gamma_{br}^\prime$, $\log\xi_s-\log\delta_b$,
$\log\xi_s-\log B^\prime$ and $\log\xi_s-\log R_b^\prime$ plots for the sample. The red point represents the source BLLAC.
}\label{figure6}
\end{figure}

In the log-log representation, $\xi_s$ as a function of $B^\prime$, $\delta_b$, $\gamma_{br}^\prime$ and $R_b^\prime$
are respectively plotted in the corresponding panel of Figure~\ref{figure6}, and the Pearson correlation coefficient and $p-$values are reported in Table \ref{rela}.
Compared with $L_{pk}^*$, the results indicate that $\xi_s$ is significantly correlated with $\gamma_{br}^\prime$, and inversely correlated with $\delta_b$.
However, $\log\xi_s$ and $\log B^\prime$ is weakly correlated with a p-value $p=0.052$ when excluding BLLAC (marked as a red point),
 while the presented correlation between $\log\xi_s$ and $\log R_b^\prime$ disappears completely. 
Omitting this object does not have a significant effect on the results in other planes.
The absence of the correlations imply that $B^\prime$ or $R_b^\prime$ may play a dominant role, while
 the presence of correlations imply that both $\delta_b$ and $\gamma_{br}^\prime$ may be rule out.

Based on the analytical theory, an inspection of the $\log\xi_s-\log\nu_{pk}^*$ plane rules out the case $\beta=\infty$ applying
if the source volume is the dominant driver of the spectral evolution. Thus, it implies that $B^\prime$ may play a dominant role.
In spite of a large $p-value$ , the resulting power-law index from our fit is $0.92$, which implies that
the spectral evolution may be instead compatible with a combined effect of variations of $B^\prime$ (corresponding $\beta=2$)
and of $\gamma_{br}^\prime$ ($\beta=0.5$).

Further, $\xi_s$ is expected to scale as $\xi_s\propto R_bB^{\prime3/2}/\delta_b^{1/2}$,
if changes of $\gamma_{br}^\prime$ dominate and $\nu_{pk}^*$ varies in a narrower range compared to $\delta_b$.
By substituting Eq.(\ref{BRR}), we can arrive $\xi_s\propto R_b^{\prime-1/14}\delta_b^{-1/2}$ or $\xi_s\propto B^{\prime1/10}\delta_b^{-1/2}$.
The resulting exponent of power-law relation between $\xi_s$ and $\delta_b$ could yield 0.5, which is much smaller than 1.8 obtained from our fit for the sample. Similarly with the analysis of the case $\beta=0.5$, $\xi_s$ is expected to scale as $\xi_s\propto \delta_b^{-2}\gamma_{br}^{\prime-2/3}$ or $\xi_s\propto\delta_b^{-2}R_{b}^{\prime-2/7}$ for the case $\beta=2.0$ applying if variations of $B^\prime$ are the dominant driver.
Thus, on the one hand the $\xi_s-\delta_b$ relation rules out the case $\beta=0.5$ applying when the spectral changes are dominated by variations of $\gamma_{br}^\prime$, on the other hand it confirms that $B^\prime$ may play the dominant role.

Although $\gamma_{br}^\prime$ is unlikely to be the dominant role, but it may be considered to be subdominant role compared to $B^\prime$.
The secondary role for $\gamma_{br}^\prime$ is confirmed by the correlation between $\xi_s$ and $\gamma_{br}^\prime$.
It can be accounted by a simple scenario in which we substitute $B^\prime\propto R_b^{\prime-0.5}$ instead of
analytical expression Eq.(\ref{BRR}) to the definition of $\xi_s$ Eq.(\ref{mag1}).

Reasonably, we conclude that $B^\prime$ may play a dominant role, and is accompanied by a rescaling of variations of $\gamma_{br}$.
Since $B^\prime$ is strongly anti-correlated with $R_b^\prime$, the dispersion of a possible relation may be mainly contributed to variations of $\delta_b$.


As a final emphasize, the correlation between $\xi_s$ and $\nu_{pk}^*$ may be in favour of a blazar sequence.
In \cite{Finke2013}, the author indicated that an anti-correlation between the Compton dominance and $\nu_{pk}^*$ may exist for BL Lac objects alone,
as no evidence for a correlation has been found for FSRQs.
In the work, the Compton dominance is defined as the ratio of the peak Compton-scattered component to the peak of the
synchrotron component, i.e., $\max[L_{pk}^{ssc},L_{pk}^{ec}]/L_{pk}^{syn}$.
For BL Lac objects, the peak Compton-scattered emission may be mainly contributed
by the SSC process, which is generally interpreted as the EC process for FSRQs.
Combining their results with ours, an inverted ``V'' shape in the $\log\nu_{pk}^*-\log\xi_s$ plot may appear,
although BL Lac objects may be out of equipartition in contrast to FSRQs \citep{Yan2014,Dermer2015}.
It may be ascribed to that the sign of the $\log\xi_s-\log\nu_{pk}^*$ relation permits us to discriminate the IC emission happening
in either the Thomson or Klein-Nishina (KN) regimes.
In the former case, $\xi_s$ increases with $\gamma_{br}^\prime$, while
it decreases as $\gamma_{br}^\prime$ increases in the transition to the KN regime \citep{TMG1998,Moderski2005}.
In fact, the factor $\xi_s$ is inversely proportional to $\gamma_{br}^{\prime2}$ when the IC scattering occurs in the deep KN regime.

\section{Summary}\label{summary}
In the present work, we modeled the quasi-simultaneous SEDs of 21 powerful blazars with a one-zone leptonic model that
considers a LPEED in a state of equipartition between the electrons and magnetic-field energy densities.
The Markov Chain Monte Carlo (MCMC) fitting procedure was employed to determine the spectral parameters,
which are consisted of the peak synchrotron luminosity ($L_{pk}^*$), its peak frequency ($\nu_{pk}^*$),
and the synchrotron self-Compton (SSC) dominance ($\xi_s$). Then, combining the spectral parameters
with the spectral curvature ($r$) of the employed EED allowed us to obtain our interesting physical quantities,
including the magnetic-field strength ($B^\prime$), Doppler factor ($\delta_b$),
radius of radiation zone ($R_b^\prime$) and the characteristic energy of EED ($\gamma_{br}^\prime$).
It is not only allow us to study the relations among the spectral parameters, but also provide us insight into the relations
between the spectral parameters and the physical properties of jets, and the relations among the relevant physical quantities.

Our results show that the employed model can give a satisfactory description of the SEDs and
the spectral break in the GeV band for the majority of the sources.
No significant correlations or trends are found in the $\log L_{pk}^*-\log\nu_{pk}^*$ and $\log\xi_s-\log\nu_{pk}^*$ planes for the sample.
However, there exists a strongly negative correlation between $B^\prime$ and $R_b^\prime$, and
our fit gives $B^\prime\propto R_b^{\prime-0.51\pm0.04}$.
The relation is according with that derived from the analytical theory,
and should be self-consistent with the underling acceleration mechanism of the employed EED.
By assuming a conical jet with opening angle $\theta_{op}\sim1/\Gamma_b$,
we find that the $\gamma$-ray emission site being located at a
distance $\sim10^{17}-10^{19}$ cm.

The absence of two possible trends in $\log\nu_{pk}^*-\log L_{pk}^*$ and $\log\nu_{pk}^*-\log\xi_s$ planes is directly related
to significant variations of physical properties of jets among the different sources.
Our analysis show that $\gamma_{br}^\prime$ and $B^\prime$ are those most relevant as dominant mechanisms.
For the spectral evolution in $\log\nu_{pk}^*-\log L_{pk}^*$ plane, the main driver should be variations of $\gamma_{br}^\prime$.
For the spectral evolution in $\log\nu_{pk}^*-\log\xi_s$ plane, we conclude that $B^\prime$ may play a dominant role, and is accompanied by a rescaling of variation of a secondary quantity,
which may be considered as $\gamma_{br}^\prime$.
The absence of a significant correlation may be primarily ascribed to changes of $\delta_b$, since $B^\prime$ and $R_b^\prime$ are anti-correlated.

\section{appendix}

\subsection{Analytical calculation of the SSC model}\label{derivations}
The emitting electrons with a LP law emit LP synchrotron and the inverse Compton emission spectra.
In general, we can consider
\begin{equation}\label{logpablic}
{n_k^\prime}(\gamma^\prime)={\gamma^\prime}^kn_e^\prime(\gamma^\prime),
\end{equation}
peaking at $\gamma_{pk}^\prime=\gamma_{br}^\prime\exp[-(s-k)/2\hat r]$ with its maximum $k_{pk}^\prime\equiv n^\prime_{k,pk}(\gamma^\prime_{pk})=k_{br}^\prime{\gamma_{br}^\prime}^k\exp[(s-k)^2/4\hat r]$, and $\hat r=r/\ln10$.
In the case $k=s$,  $\gamma_{pk}^\prime=\gamma_{br}^\prime$ and $n^\prime_{e,pk}(\gamma^\prime_{pk})=k_{br}^\prime{\gamma_{br}^\prime}^k$.
In turn, inverting $\gamma_{pk}^\prime$ and $k_{pk}^\prime$, and is substituted into Eq.\ref{logpablic}, we obtain an equivalent functional relation, and is given by
\begin{equation}
{n_k^\prime}(\gamma^\prime)=k_{pk}^\prime\Big(\frac{\gamma^\prime}{\gamma_{pk}^\prime}\Big)^{-\hat{r}\ln(\gamma^\prime/\gamma_{pk}^\prime)}.
\end{equation}
This indicate that the function ${n_k^\prime}(\gamma^\prime)$ is only determined by three free parameters including the curvature $r$, peak Lorentz factor $\gamma_{pk}^\prime$ and its maximum $k_{pk}^\prime$. 
Taking the derivatives of Eq.\ref{logpablic} with respect to $\gamma^\prime$, we obtain
\begin{equation}
\frac{dn_k^\prime}{d\gamma^\prime}=\frac{n_k^\prime d\ln n_k^\prime}{\gamma^\prime d\ln(\gamma^\prime/\gamma_{br})^\prime}
=\frac{n^\prime_{k}(\gamma^\prime)}{\gamma^\prime}(k-s-2\hat{r}\ln\frac{\gamma^\prime}{\gamma_{br}^\prime}).
\end{equation}
It shows that the derivatives of a LP function is still close to a LP law.

Integrating Eq.\ref{logpablic} over $\gamma^\prime$ gives
\begin{equation}\label{inte1}
\int_0^\infty{\gamma^\prime}^kn_e^\prime(\gamma^\prime)d\gamma^\prime=k_{br}^\prime{\gamma_{br}^\prime}^{k+1}\exp\Big\{\frac{(s-k-1)^2}{4\hat r}\Big\}\sqrt{\frac{\pi}{\hat r}}.
\end{equation}

From above equation, the various momenta of the distribution is obtained by
\begin{equation}
\langle{\gamma^\prime}^k\rangle=\frac{\int{\gamma^\prime}^k n_e(\gamma^\prime)d\gamma^\prime}{\int n_e(\gamma^\prime)d\gamma^\prime}={\gamma_{br}^\prime}^k\exp\{[k^2-2k(s-1)]/4\hat r\}.
\end{equation}
Note that the root mean square (r.m.s) energy satisfies $\sqrt{\langle{\gamma^\prime}^2\rangle}=\gamma_{pk}^\prime$.

The energy density of the non-thermal electrons in the co-moving frame is
\begin{eqnarray}
u_e^\prime &=& m_ec^2\int_1^\infty d\gamma^\prime\gamma^\prime  n_e^\prime(\gamma^\prime)\nonumber\\
&=&k_{br}^\prime{\gamma_{br}^\prime}^2m_ec^2\int_{1/\gamma_{br}^\prime}^\infty dxx^{1-s-\hat{r}\ln{x}}.
\end{eqnarray}
If we replace $1/\gamma_{br}^\prime$ with 0, and $u_e^\prime=\xi_eU_B^\prime$, the normalization factor of relativistic electron spectrum can be written in the form of
\begin{equation}\label{elenorm}
k_{br}^\prime\simeq\frac{\xi_eU_B^\prime}{{\gamma_{br}^\prime}^2m_ec^2I_1},
\end{equation}
where the function $I_1=f_0f_1$, $f_1=\exp[(s-2)^2/4\hat r]$, $f_0=\sqrt{\pi/\hat r}$.

Using a $\delta$-function approximation, we can obtained the luminosity at a specific frequency in the blob's reference frame,
\begin{equation}\label{deltasyn}
\epsilon^\prime L_{syn}^\prime(\epsilon^\prime) = |\dot\gamma^\prime|m_ec^2V_b^\prime \gamma^\prime n_e^\prime(\gamma^\prime)[d\ln\gamma^\prime/d\ln\epsilon^\prime].
\end{equation}
where $\epsilon^\prime$ is related to $\gamma^\prime$ through the relation $\epsilon^\prime=(3/2){\gamma^\prime}^2\epsilon_B^\prime$,
and the synchrotron cooling rate is $\dot\gamma^\prime=-(4/3)c\sigma_TU_B^\prime\gamma^{\prime2}/m_ec^2$,
with $U_B=\epsilon_B^{\prime2}B_{cr}/8\pi$.
Here, $\epsilon_B^\prime\equiv B^\prime/B_{cr}$ is the ratio of magnetic strength and the critical magnetic field $B_{cr}=m_e^2c^3/e\hbar$,
$\sigma_T$ is the Thomson cross section, and $\hbar$ denotes the Planck's constant divided by $2\pi$.
The total synchrotron luminosity of the electrons can be written as
\begin{eqnarray}
  L_{syn} &=& m_ec^2 V_b^\prime\delta_b^4\int_1^\infty|\dot\gamma^\prime|n_e^\prime(\gamma^\prime)d\gamma^\prime \\\nonumber
   &=& \frac{4}{3}c\sigma_TU_B^\prime V_b^\prime k_{br}^\prime{\gamma_{br}^\prime}^3\delta_b^4\int_{1/\gamma_{br}^\prime}^\infty dxx^{2-s-\hat{r}\ln{x}}
\end{eqnarray}
in which $x=\gamma^\prime/\gamma_{br}^\prime$ and the synchrotron self-absorption corrections is not taken into account. Substituting Eq.\ref{elenorm} into above equation, and using
the integral formula Eq.\ref{inte1}, one can derive
\begin{equation}\label{eq1}
L_{syn}\simeq\frac{4c\sigma_TI_2}{3m_ec^2I_1}\xi_e V_b^\prime{\gamma_{br}^\prime}U_B^{\prime2}\delta_b^4
\end{equation}
where $I_2=f_0f_2$, $f_2=\exp[(s-3)^2/4\hat r]$

Then, the corresponding luminosity in the observer's frame is given by
\begin{equation}\label{lumino}
\epsilon L_{syn}(\epsilon)=\frac{2c\sigma_T}{3m_ec^2I_1}V_b^\prime\gamma^\prime_{br}\xi_e{U_B^\prime}^2\delta_b^4 x^{3-s-\hat r\ln{x}}
\end{equation}
where $x=\sqrt{\epsilon/\epsilon_{br}}$, the relation, $\epsilon=\delta_b\epsilon^\prime/(1+z)$, relates the observed photons energy $\epsilon m_ec^2$ to co-moving photon energy $\epsilon^\prime m_ec^2$, and $\epsilon_{br}^\prime=(3/2)\epsilon_B^\prime{\gamma^\prime_{br}}^2$.
The synchrotron luminosity reaches its maximum
\begin{equation}\label{eq3}
\epsilon_{s,pk} L^{syn}_{s,pk}=\frac{c\sigma_T\xi_e V_b^\prime{\gamma_{br}^\prime}}{96\pi^2m_ec^2}\frac{f_2}{I_1}{B^\prime}^4\delta_b^4,
\end{equation}
at the dimensionless peak energy,
\begin{equation}\label{eq2}
\epsilon_{s,pk}=\frac{3}{2}\frac{{\gamma_{br}^\prime}^2}{B_{cr}}\frac{f_3^2}{1+z}B^\prime\delta_b,
\end{equation}
which corresponds to the electron energy $\gamma_{pk}^\prime=\gamma_{br}^\prime f_3$ and $f_3=\exp[(3-s)/2\hat r]$.

Comparison of Eq.\ref{eq1} with Eq.\ref{eq3},
the total synchrotron luminosity $L_{syn}$ and the peak synchrotron luminosity $\epsilon_{s,p} L^{syn}_{s,p}$ is related by $L_{syn}=2f_0\epsilon_{s,pk}L^{syn}_{s,pk}$.
Substituting Eq.\ref{eq1} into $u_{syn}^\prime=L_{syn}/4\pi{R_b^{\prime2}}cf_g\delta_b^4$ for the average synchrotron energy density, one can obtained the following expression for the
the ratio between internal synchrotron and magnetic field energy densities in the comoving frame:
\begin{equation}\label{magnetic}
\xi_s=\frac{\sigma_TR_b^\prime}{18\pi m_ec^2}\frac{f_2\xi_e}{f_1f_g}\gamma_{br}^\prime B^{\prime2}.
\end{equation}

Substituting $\gamma_{br}^\prime$ derived from Eq.\ref{eq2} into Eq.\ref{eq3} and Eq.\ref{magnetic}, respectively, we obtain
\begin{equation}\label{BD1}
B^{\prime7}\delta_b^{13}=c_1[\epsilon_{s,pk}L_{s,pk}^{syn}]^2\frac{I_1^2f_3^2(1+z)^5}{\epsilon_{s,pk}\xi_e^2f_2^2t^6_{var}},
\end{equation}
\begin{equation}\label{BD2}
B^{\prime3}\delta_b=c_2\frac{f_1^2f_3^2f_g^2\xi_s^2}{f_2^2\xi_e^2t_{var}^2}\frac{1+z}{\epsilon_{s,pk}},
\end{equation}
where $c_1=\frac{3}{2B_{cr}}\Big(\frac{72\pi m_e}{c^2\sigma_T}\Big)^2$, and $c_2=\frac{3}{2B_{cr}}\Big(\frac{18\pi m_ec^2}{c\sigma_T}\Big)^2$.

Combination of Eq.\ref{eq2}, Eq.\ref{BD1} and Eq.\ref{BD2}, we have
\begin{eqnarray}
B^\prime&\simeq &11.48\frac{(1+z)^{1/4}\xi_s^{13/16}f_1^{13/16}f_3^{3/4}f_g^{13/16}}{L_{48}^{1/16}\nu_{13}^{3/8}t_4^{5/8}\xi_e^{3/4}I_1^{1/16}f_2^{3/4}},\label{analy1}\\
\delta_b& \simeq &14.97\frac{(1+z)^{1/4}L_{48}^{3/16}\nu_{13}^{1/8}\xi_e^{1/4}f_2^{1/4}I_1^{3/16}}{t_4^{1/8}\xi_s^{7/16}f_1^{7/16}f_3^{1/4}f_g^{7/16}},\label{analy2}\\
\gamma_{br}^\prime &\simeq &117.73\frac{(1+z)^{1/4}\nu_{13}^{5/8}t_4^{3/8}\xi_e^{1/4}f_2^{1/4}}{L_{48}^{1/16}\xi_s^{3/16}f_1^{3/16}I_1^{1/16}f_3^{5/4}f_g^{3/16}},\label{analy3}\\
R_b^\prime &\simeq &4.48\times10^{15}\frac{t_4^{7/8}L_{48}^{3/16}\nu_{13}^{1/8}\xi_e^{1/4}f_2^{1/4}I_1^{3/16}}{(1+z)^{3/4}\xi_s^{7/16}f_1^{7/16}f_3^{1/4}f_g^{7/16}}.\label{analy4}
\end{eqnarray}
Here the intrinsic size $R_b^\prime$ of the blob is directly connected to the minimum timescale of variation, $t_{var}$, through the causality relation $R_b^\prime=ct_{var}\delta_b/(1+z)$. 

\subsection{Spectral Energy Distributions}
\begin{figure*}[t]
 \centering
 \includegraphics[width=7.0cm]{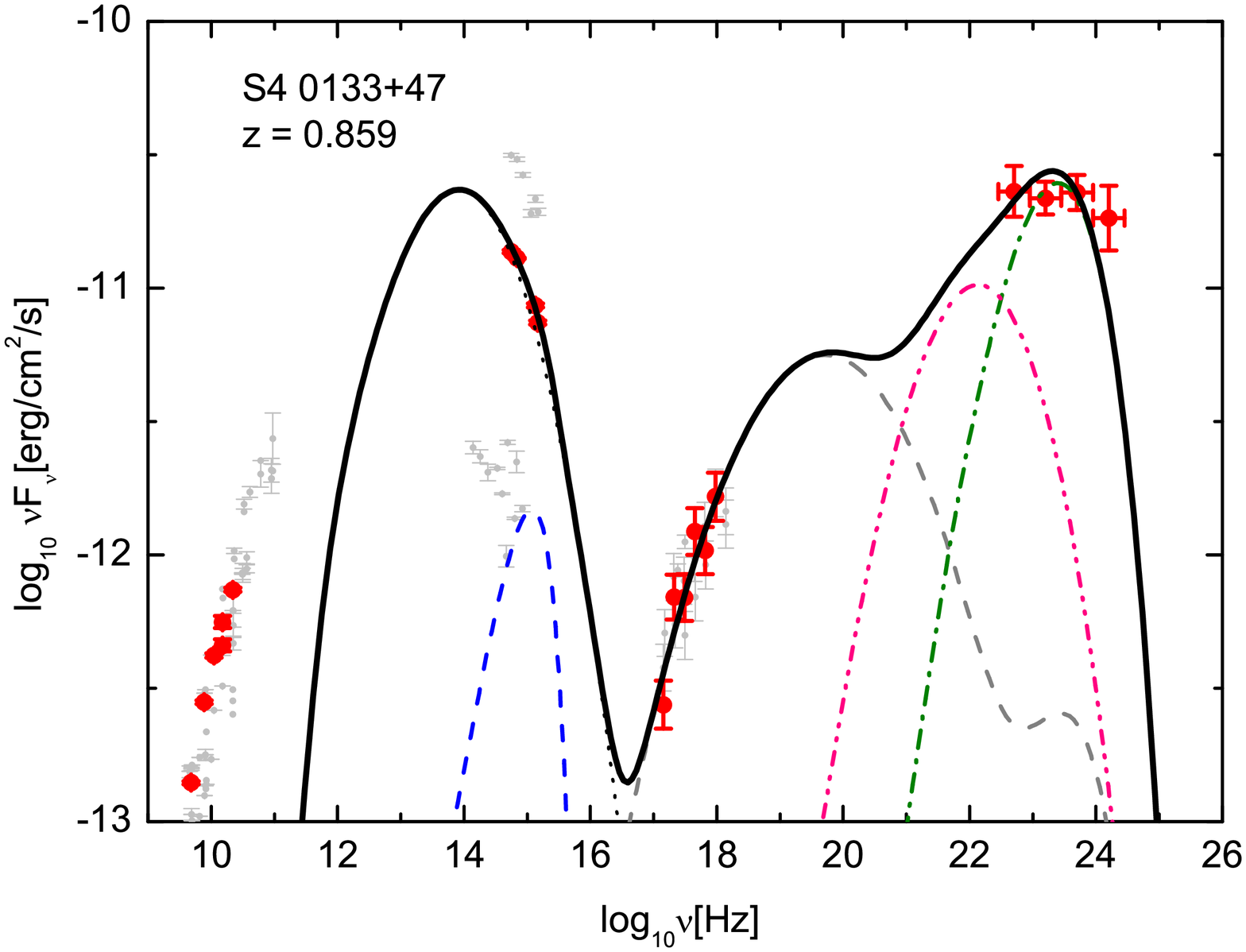}
 \includegraphics[width=7.0cm]{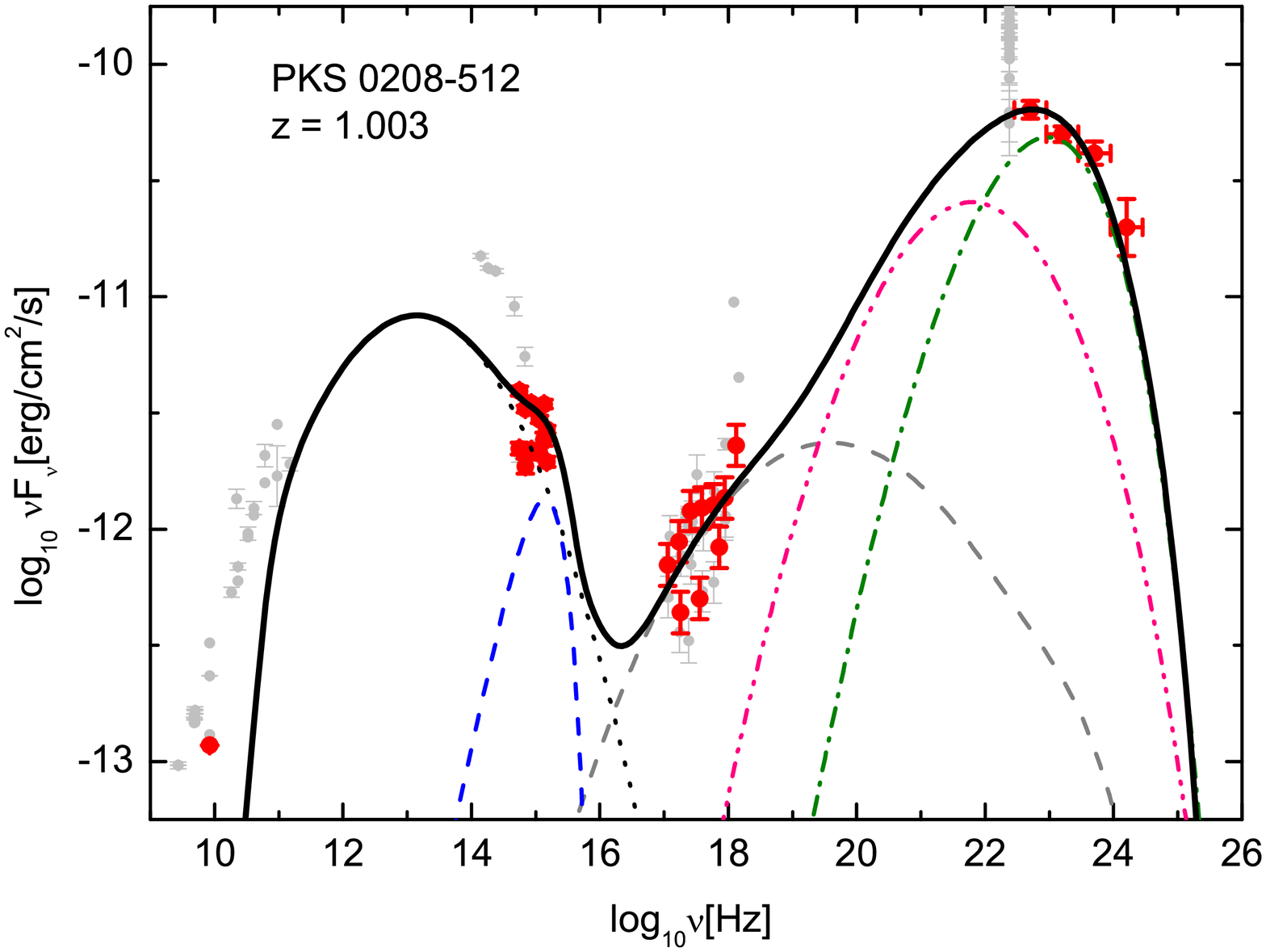}
 \includegraphics[width=7.0cm]{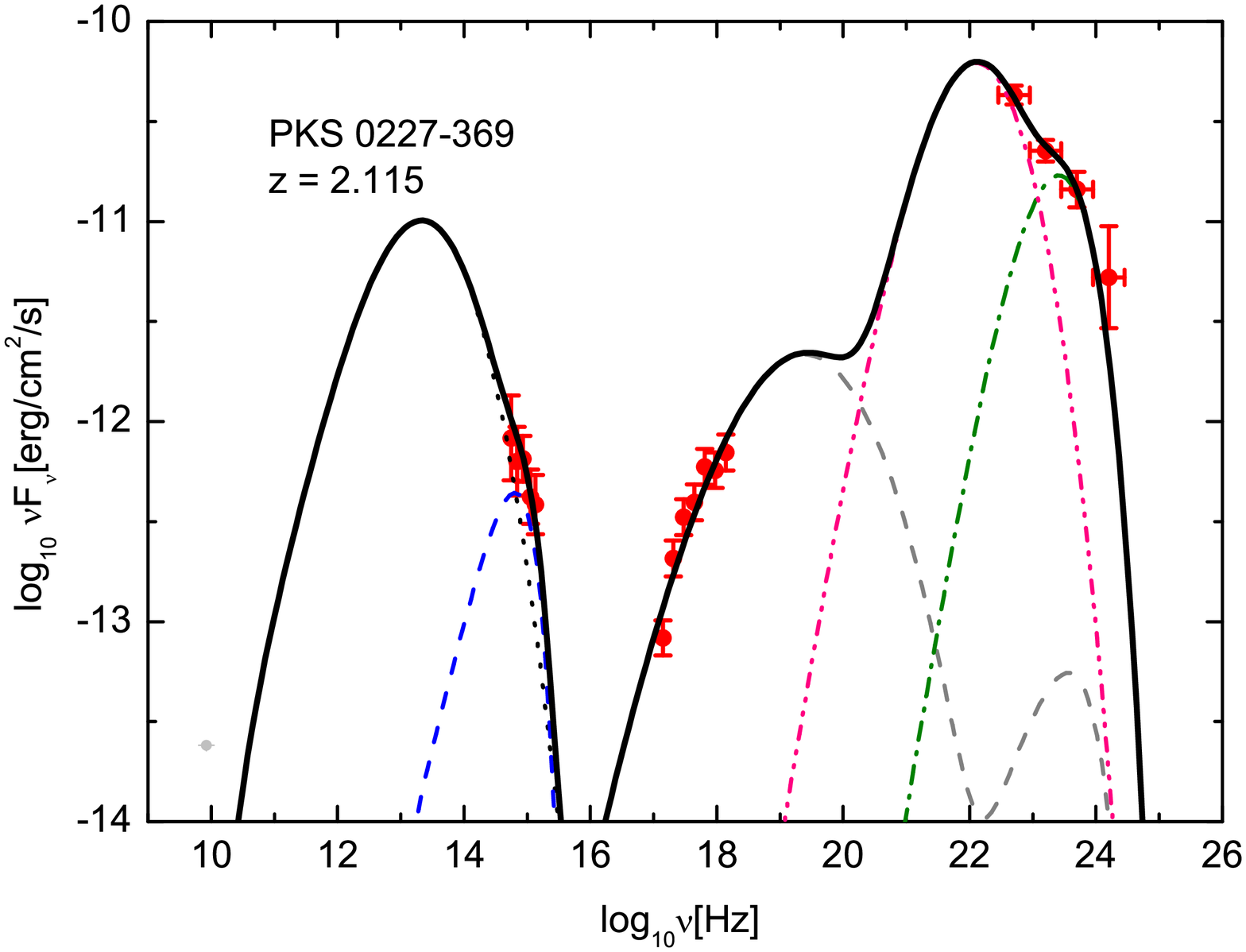}
 \includegraphics[width=7.0cm]{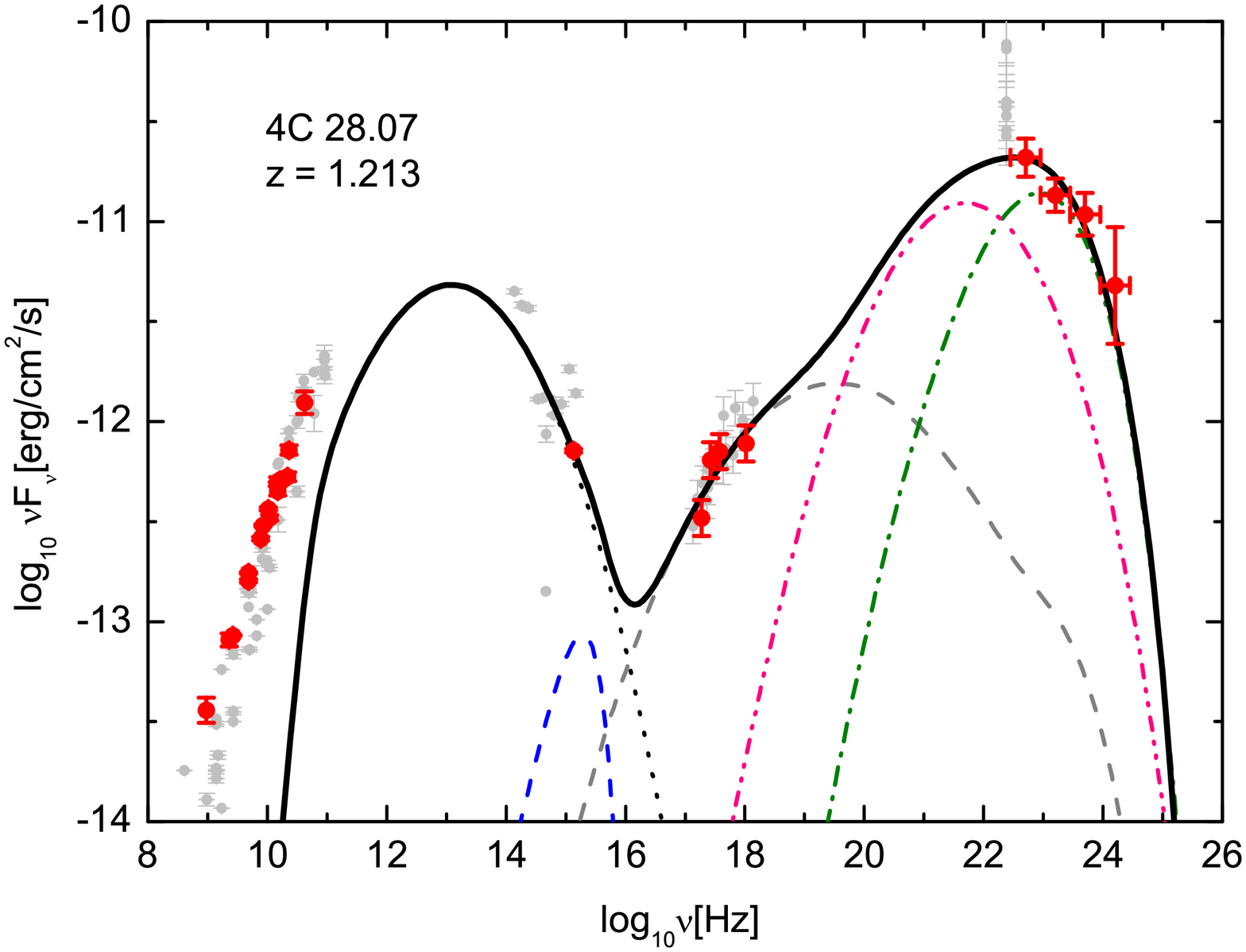}
 \includegraphics[width=7.0cm]{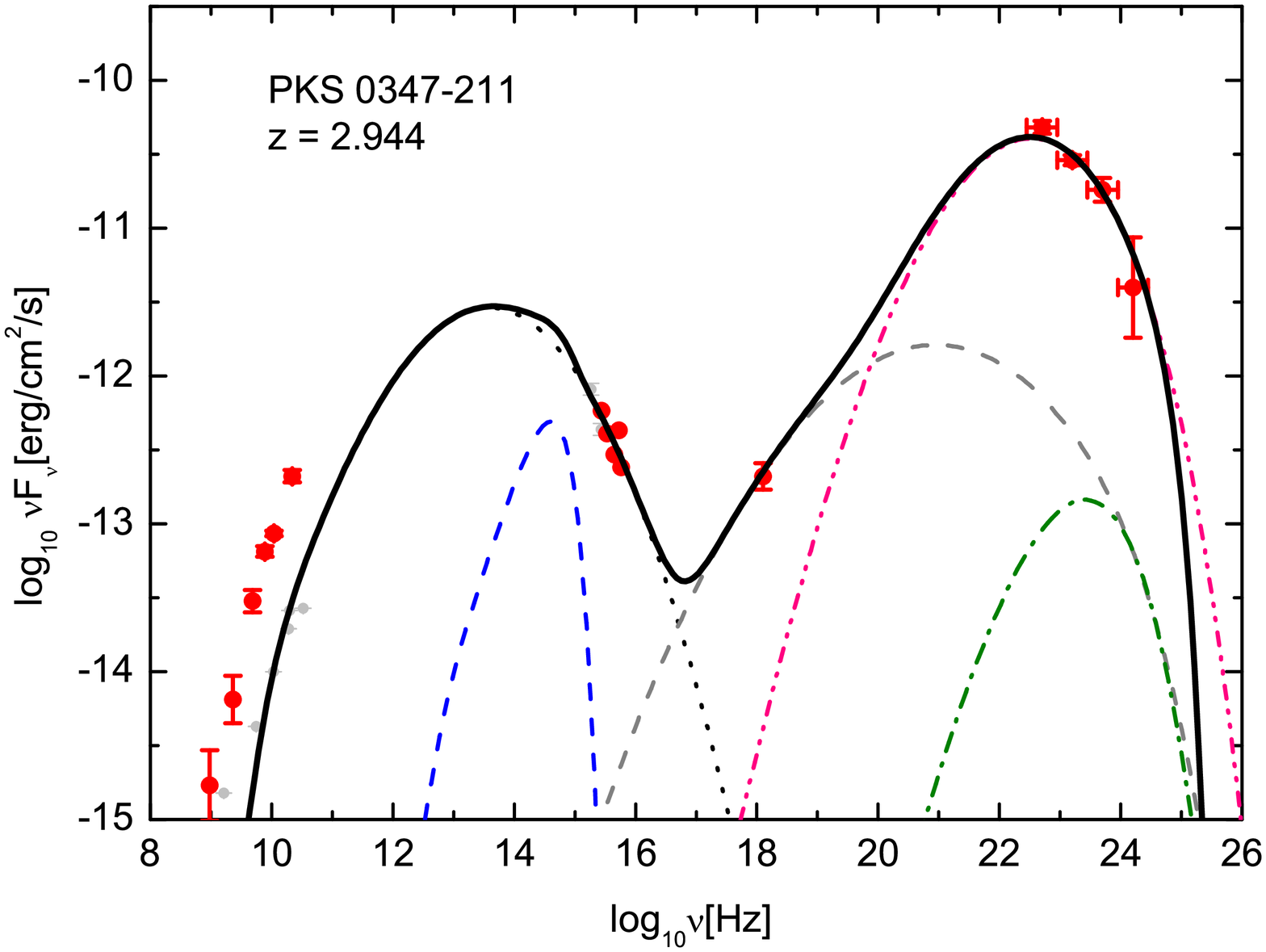}
 \includegraphics[width=7.0cm]{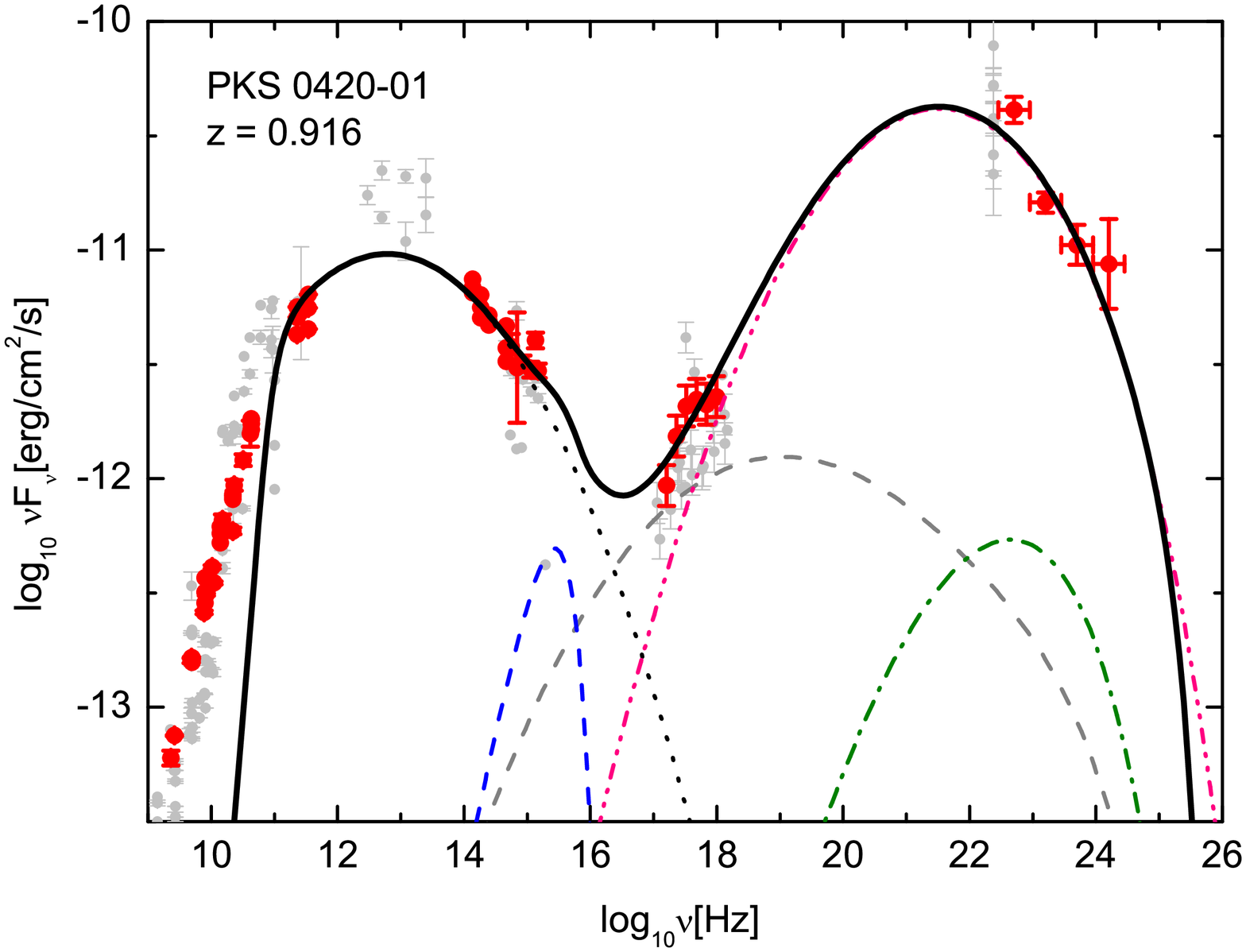}
 \includegraphics[width=7.0cm]{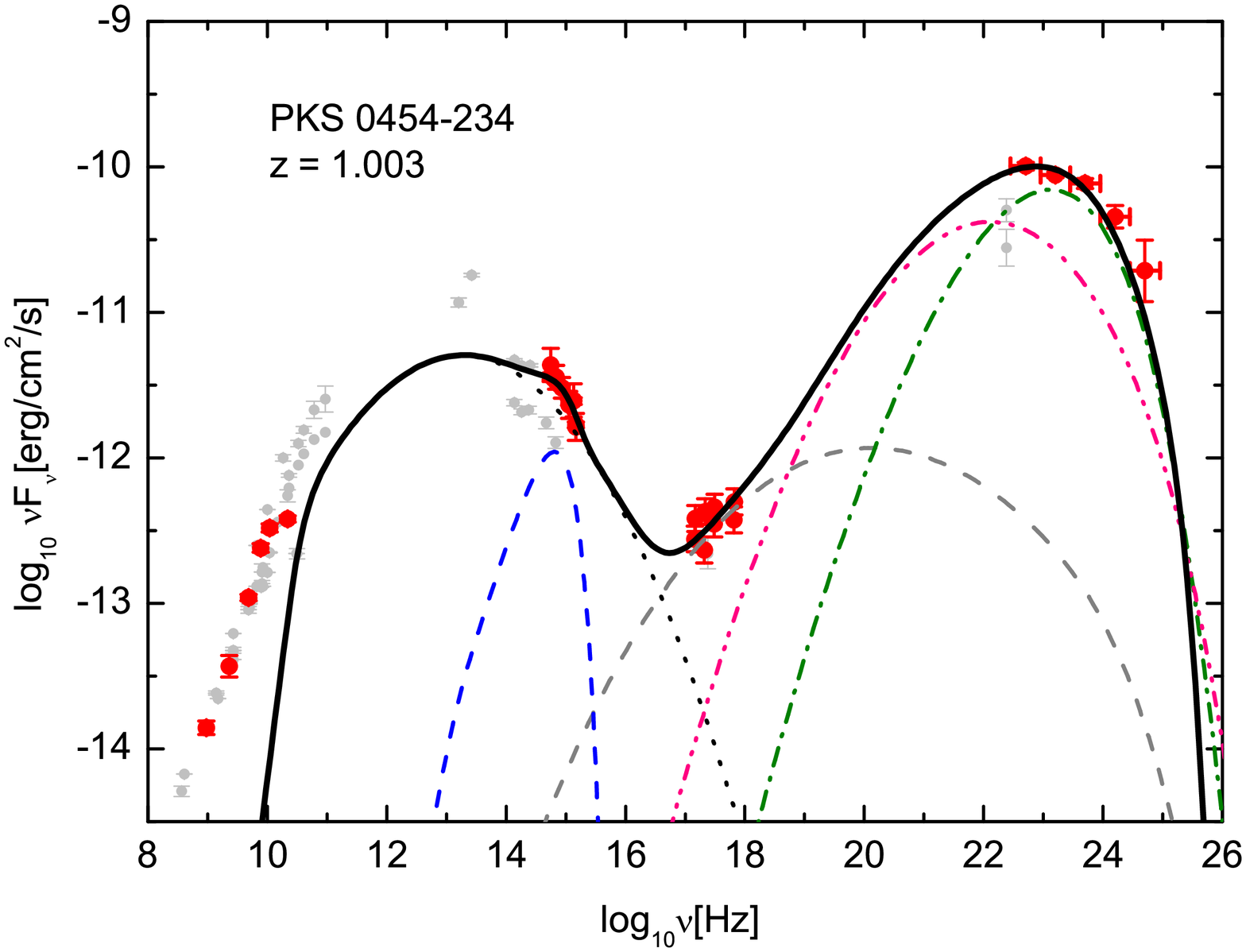}
 \includegraphics[width=7.0cm]{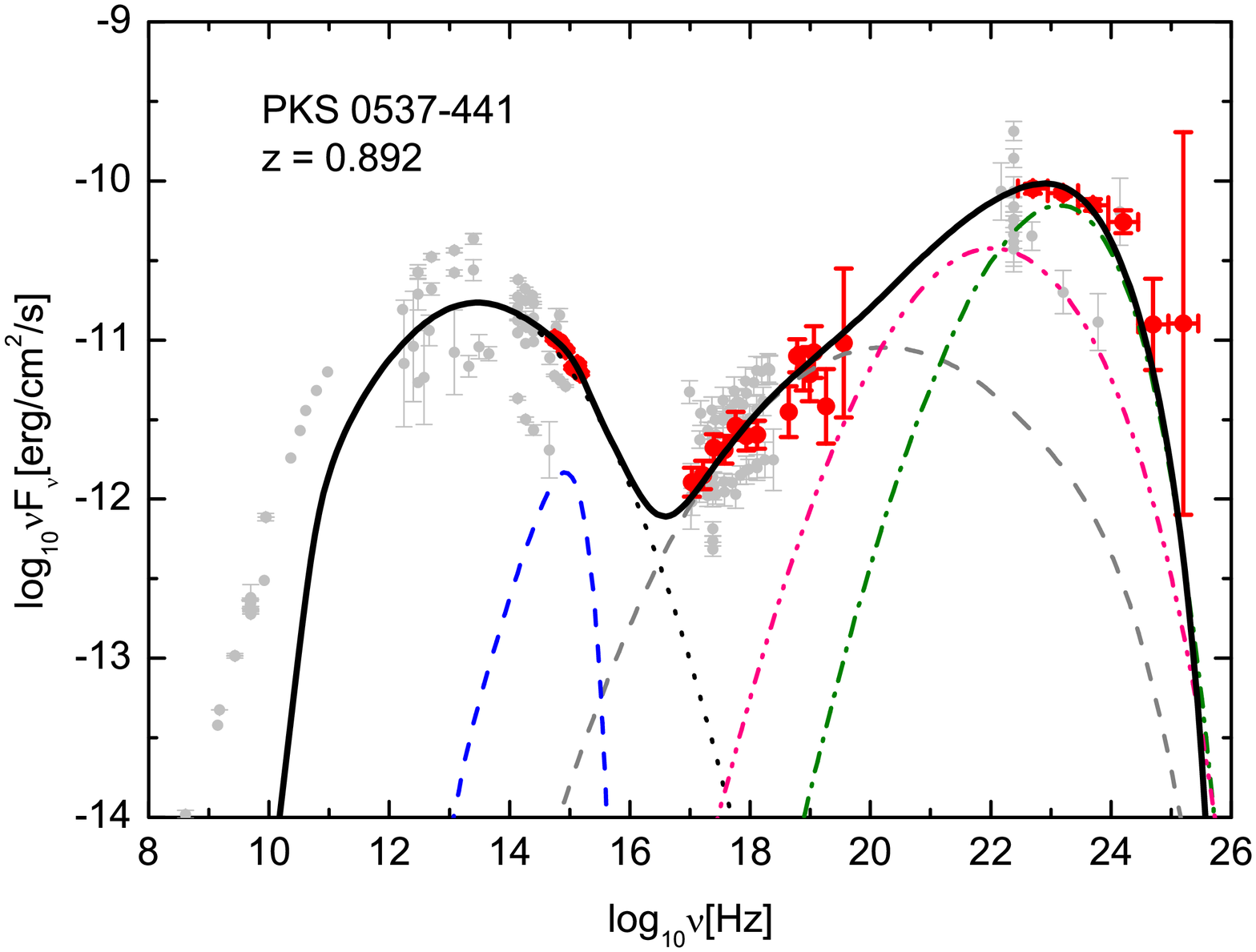}
 \caption{Comparisons of the modelling SEDs with the observed data. For each source,
 the dotted, grey dashed, dotted-dashed and the pink dotted-dotted-dashed lines are the synchrotron, SSC, external Compton-scattered BLR and DT radiation, respectively.
 The thick solid line is the sum of the all components. The quasi-simultaneous MWL data
  are indicated by solid symbols (red in the electronic version) while archival data are in light grey.}
 \label{figsed1}
\end{figure*}

\begin{figure*}[t]
 \centering
 \includegraphics[width=7.0cm]{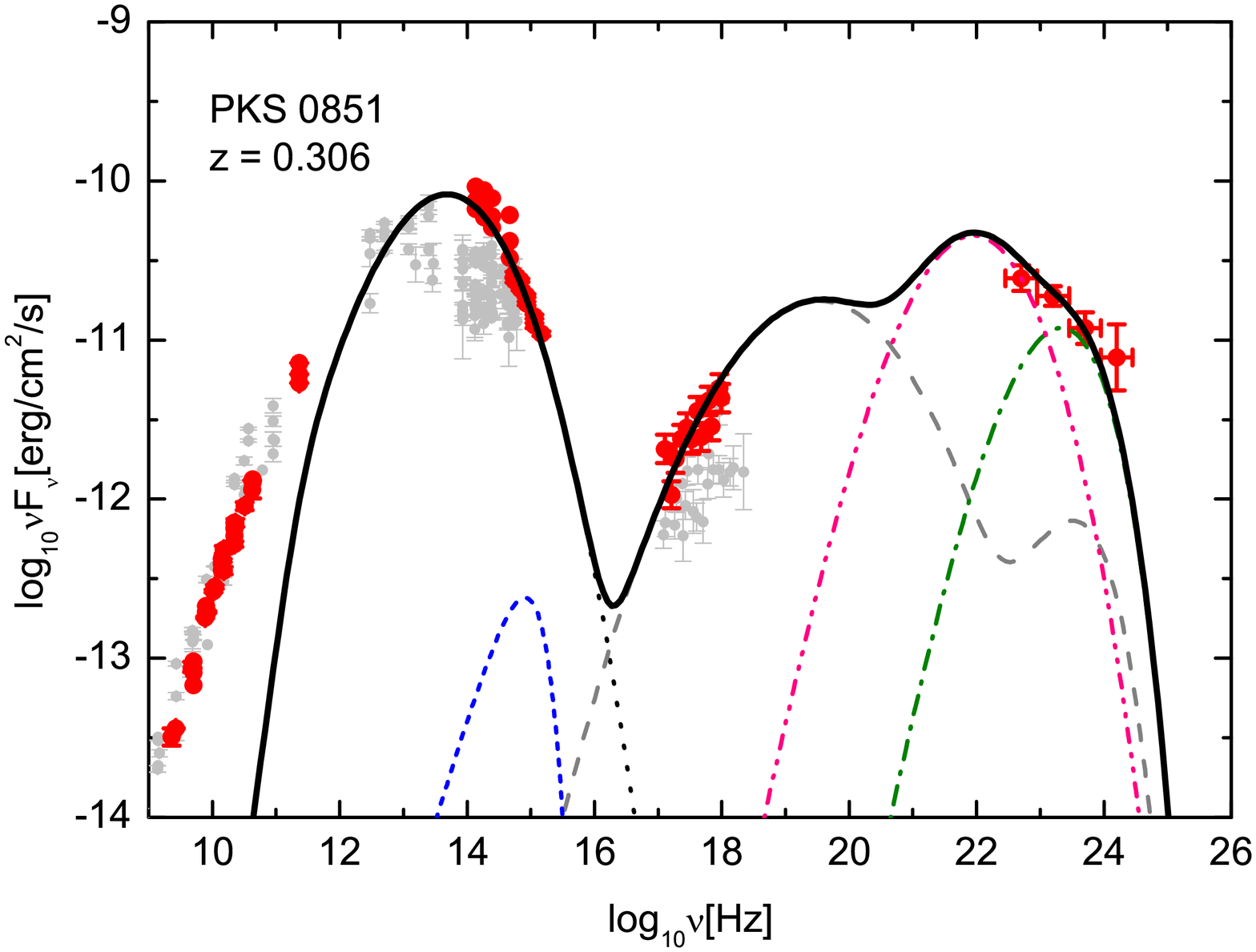}
 \includegraphics[width=7.0cm]{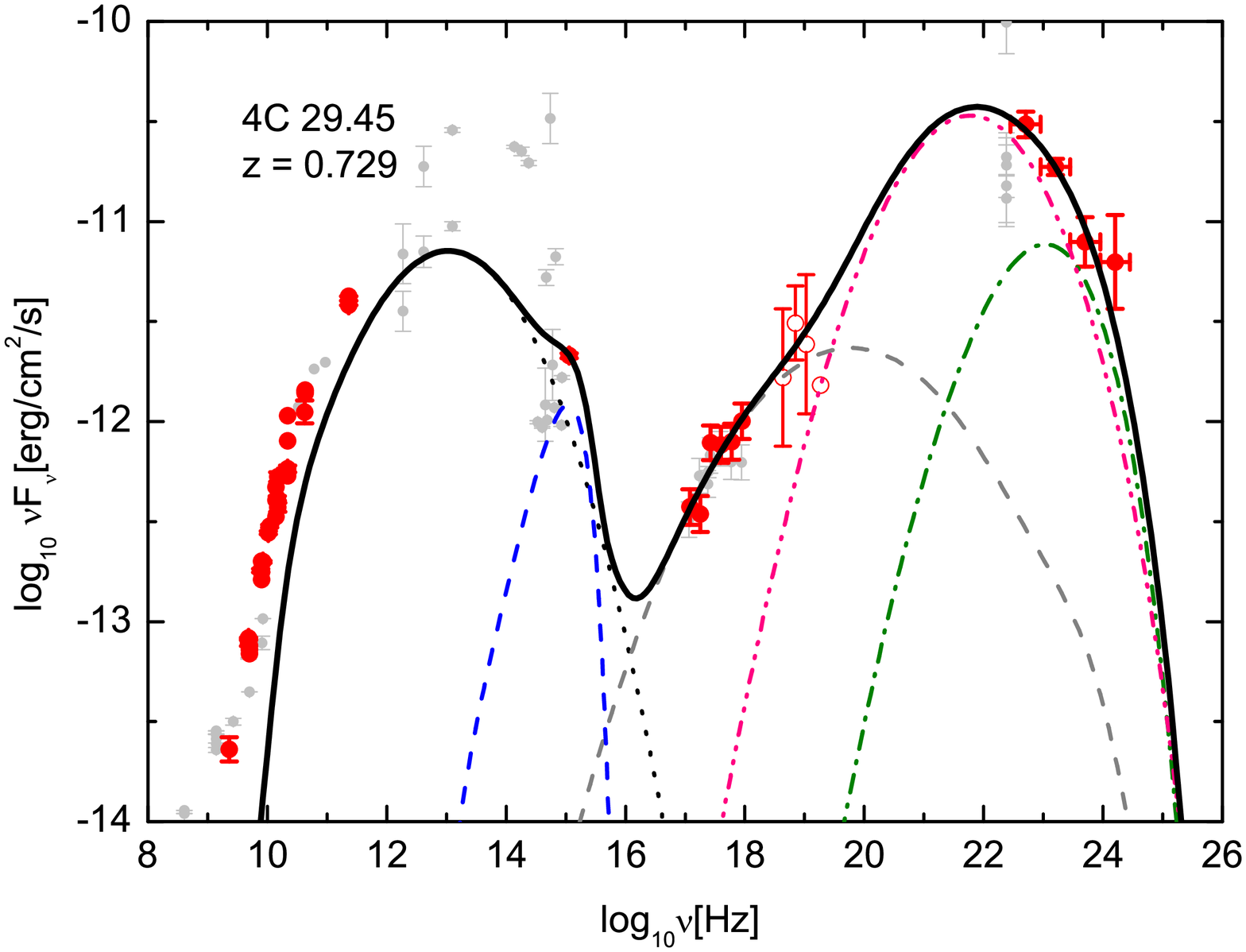}
 \includegraphics[width=7.0cm]{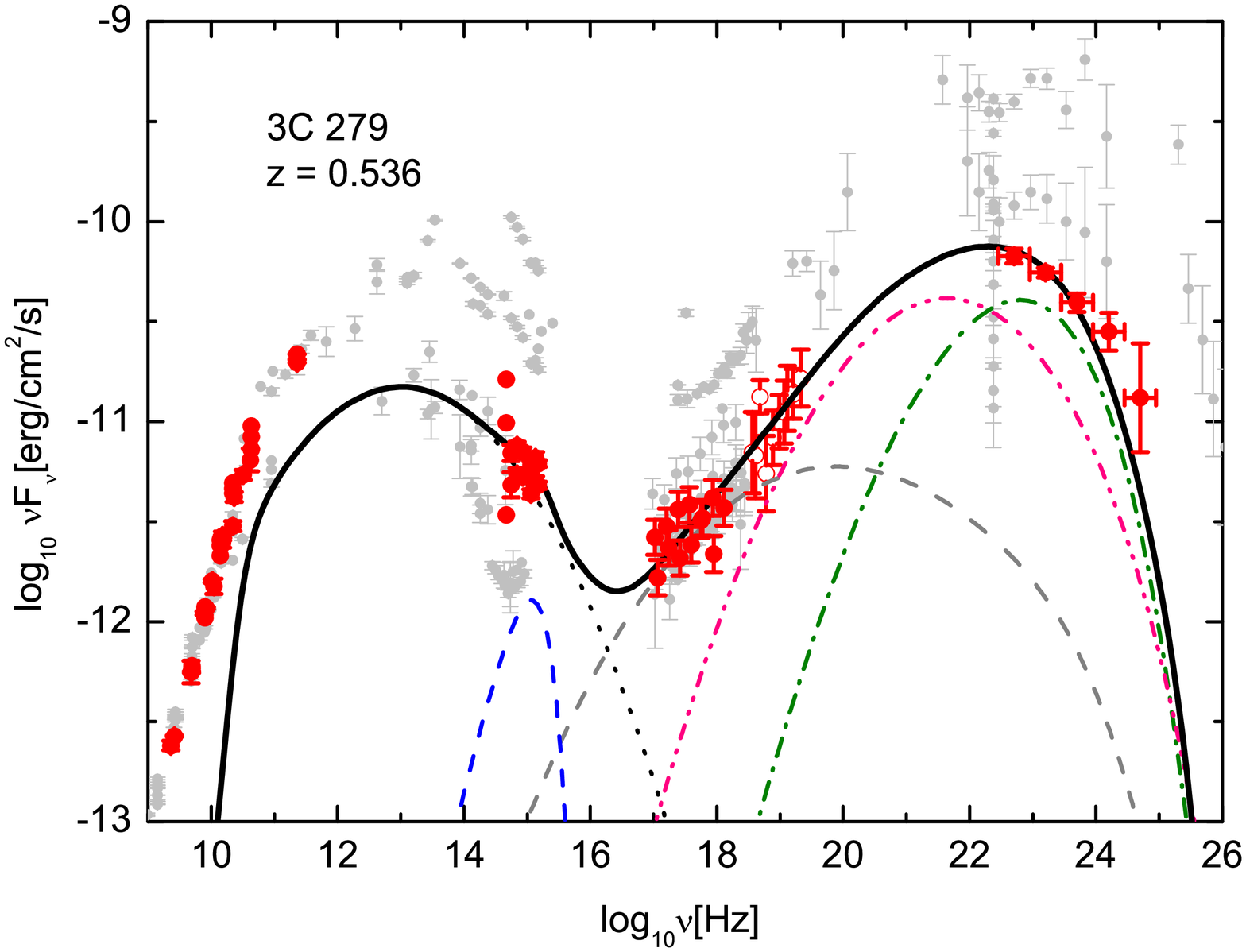}
 \includegraphics[width=7.0cm]{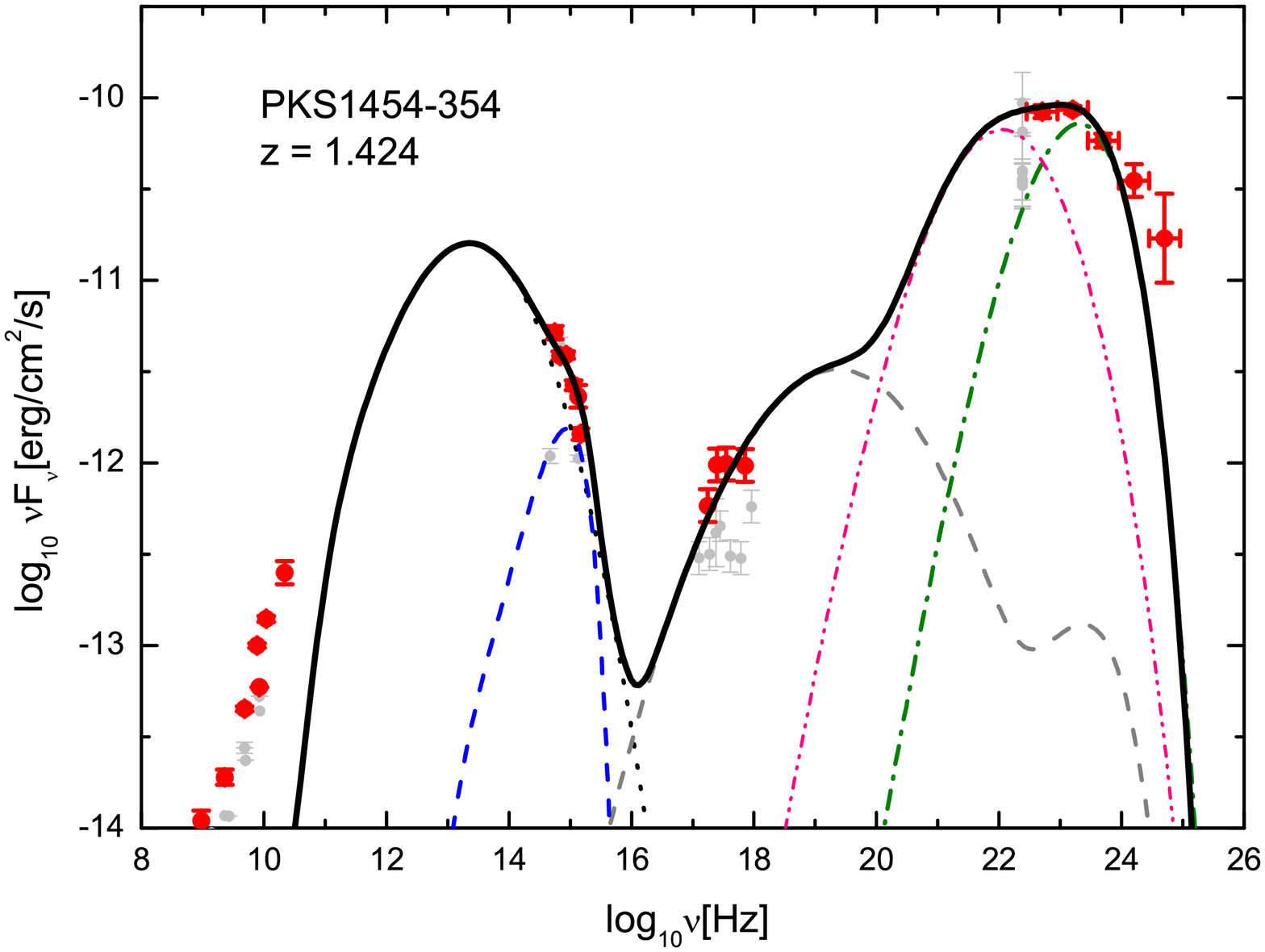}
 \includegraphics[width=7.0cm]{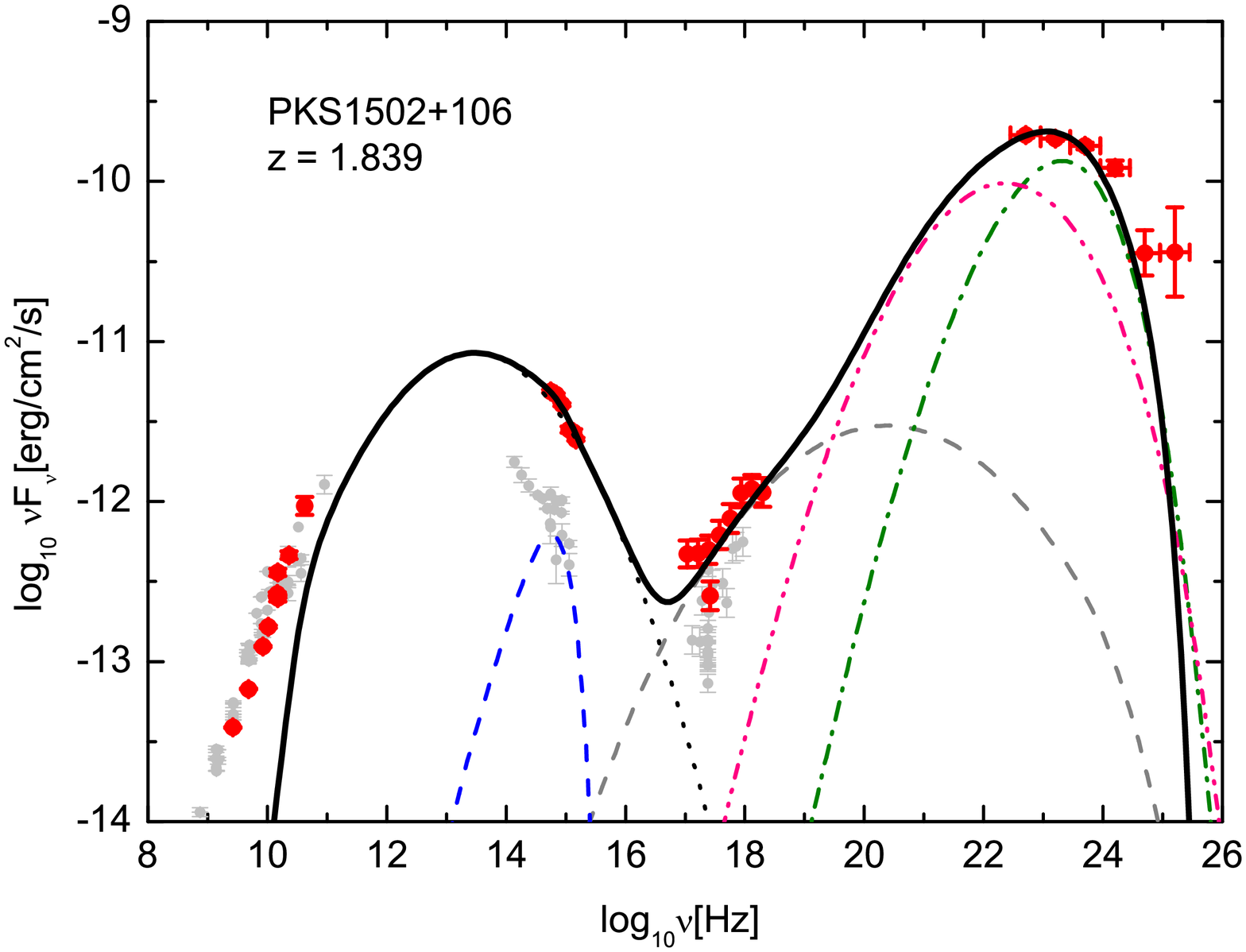}
 \includegraphics[width=7.0cm]{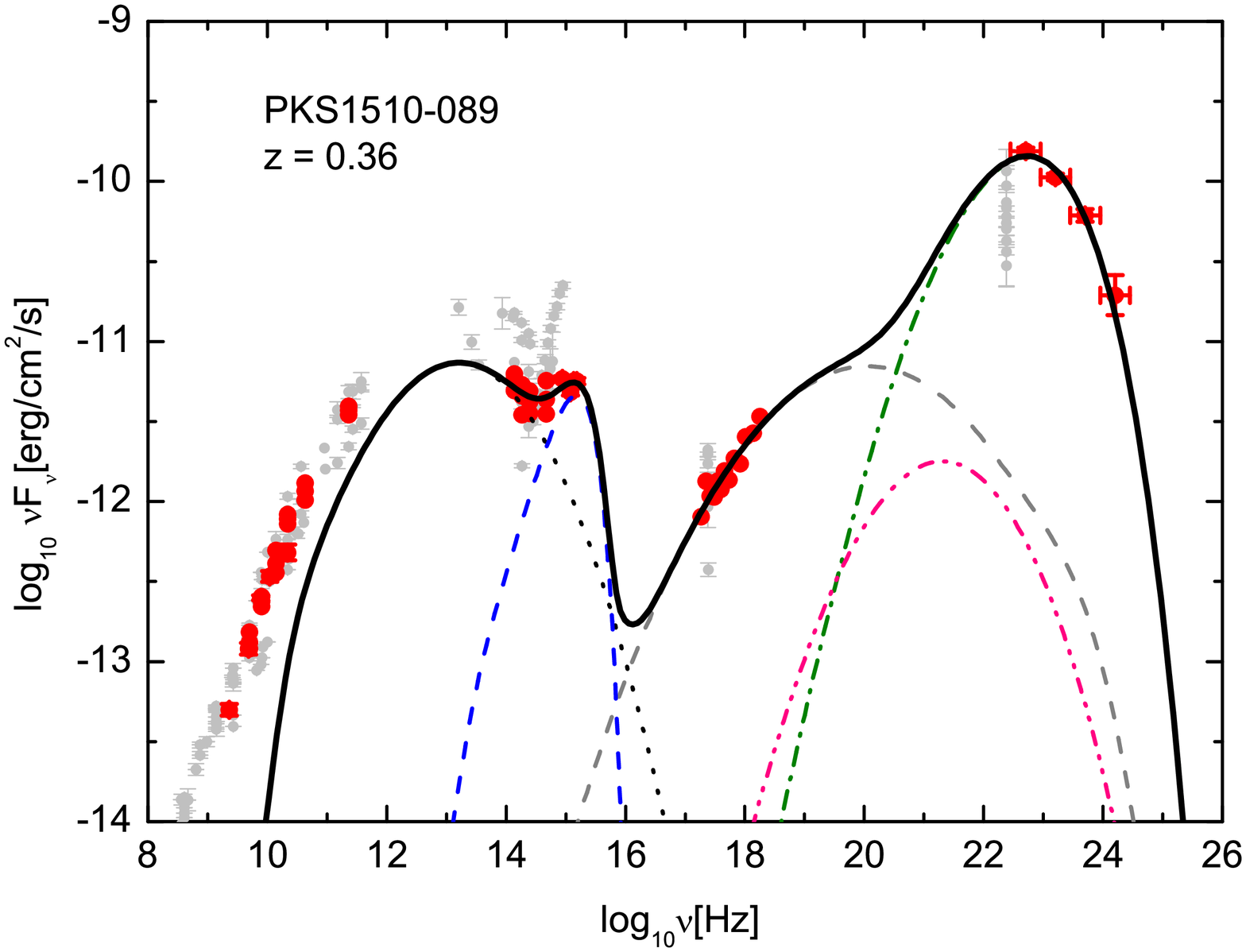}
 \includegraphics[width=7.0cm]{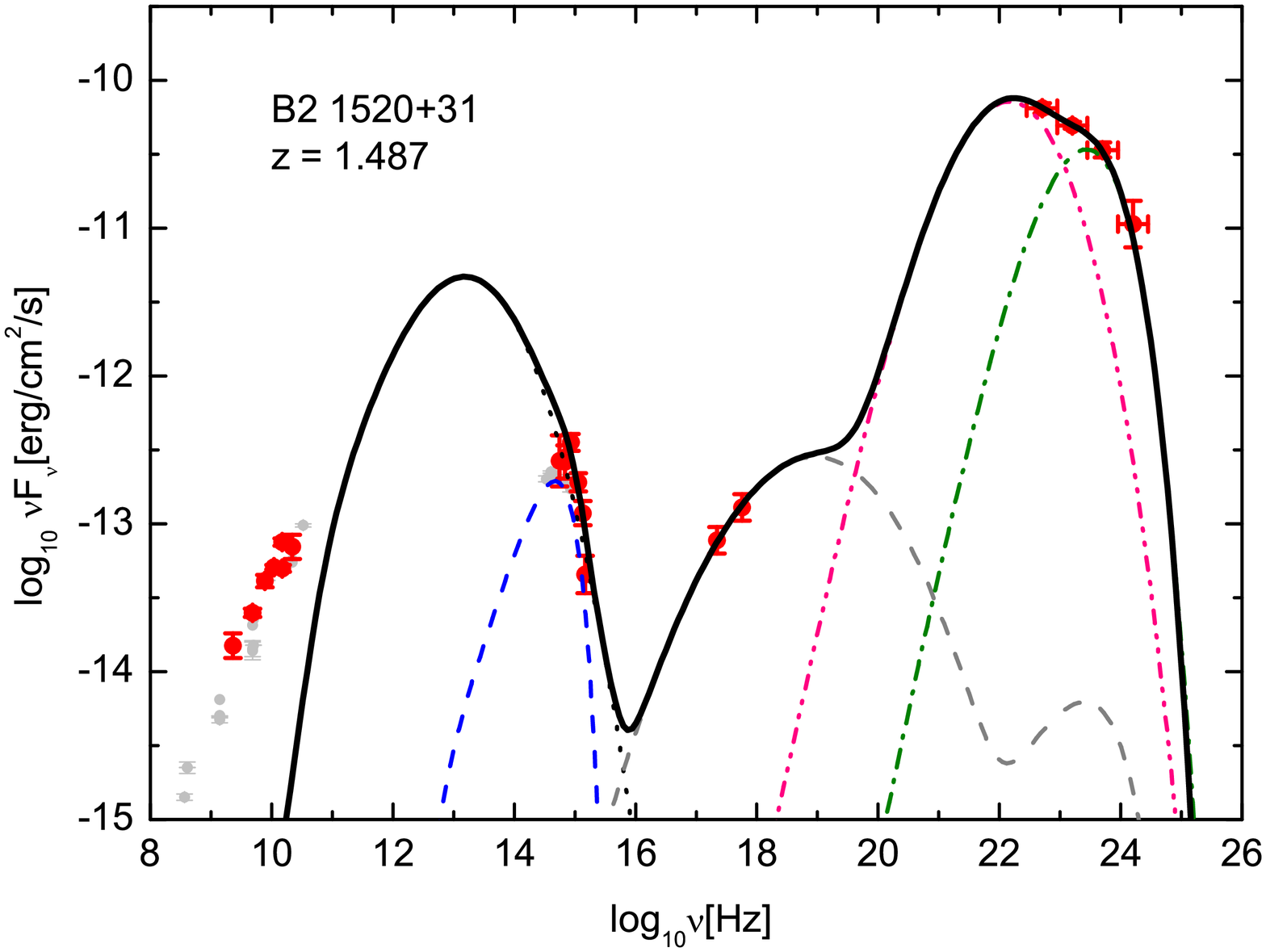}
 \includegraphics[width=7.0cm]{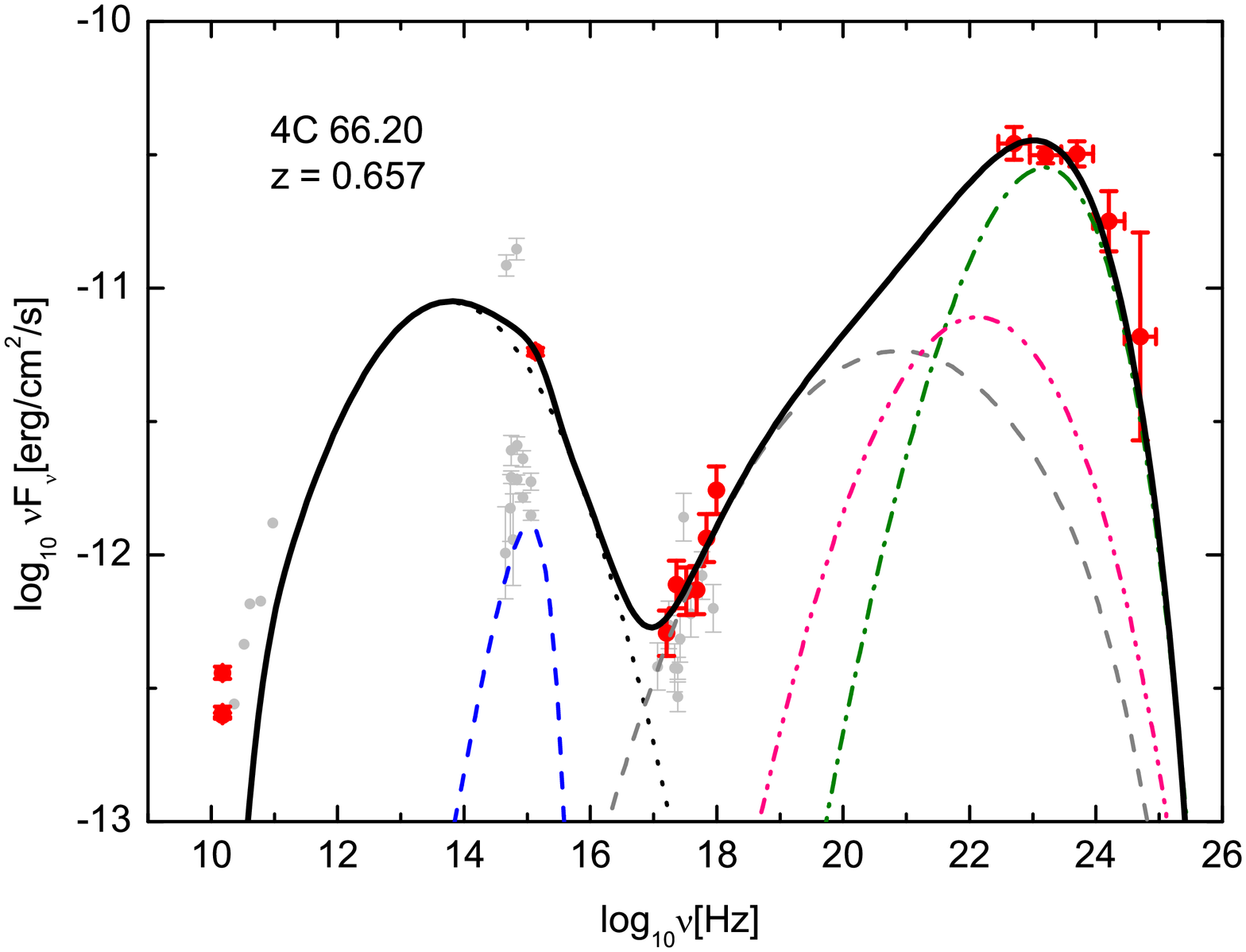}
 \caption{Same as Figure~\ref{figsed1}, but for different sources.}\label{figsed2}
\end{figure*}

\begin{figure*}[t]
 \centering
 \includegraphics[width=7.0cm]{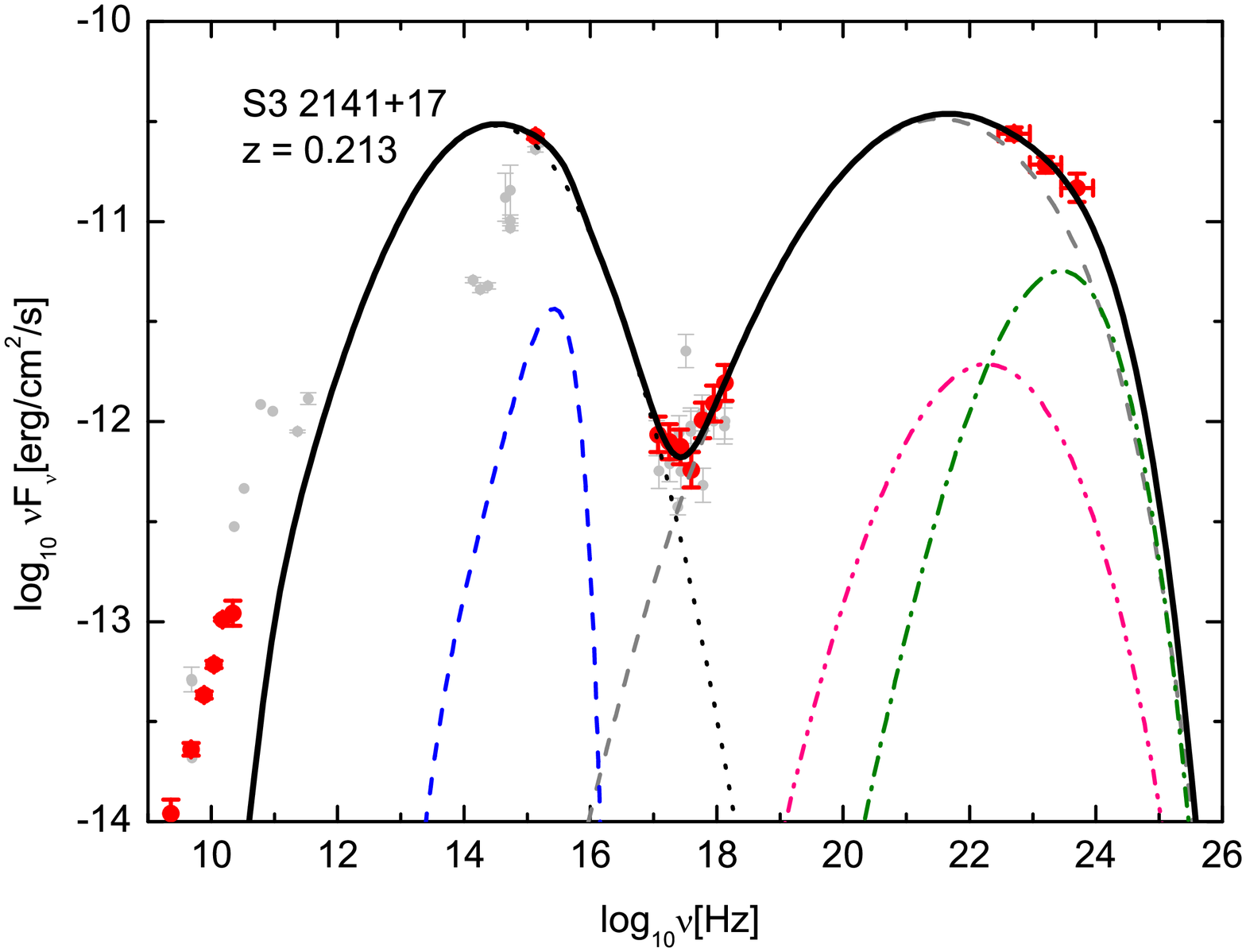}
 \includegraphics[width=7.0cm]{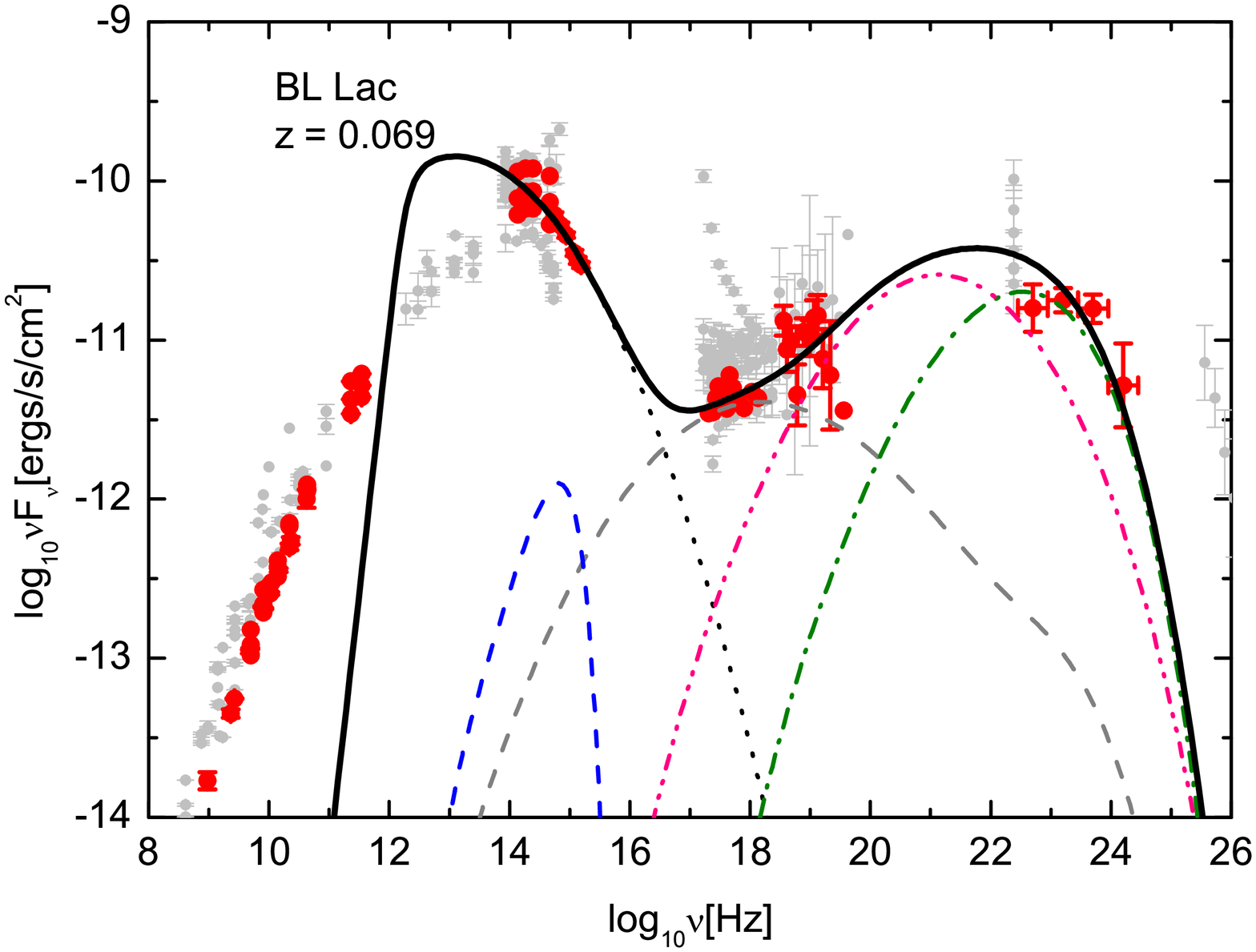}
 \includegraphics[width=7.0cm]{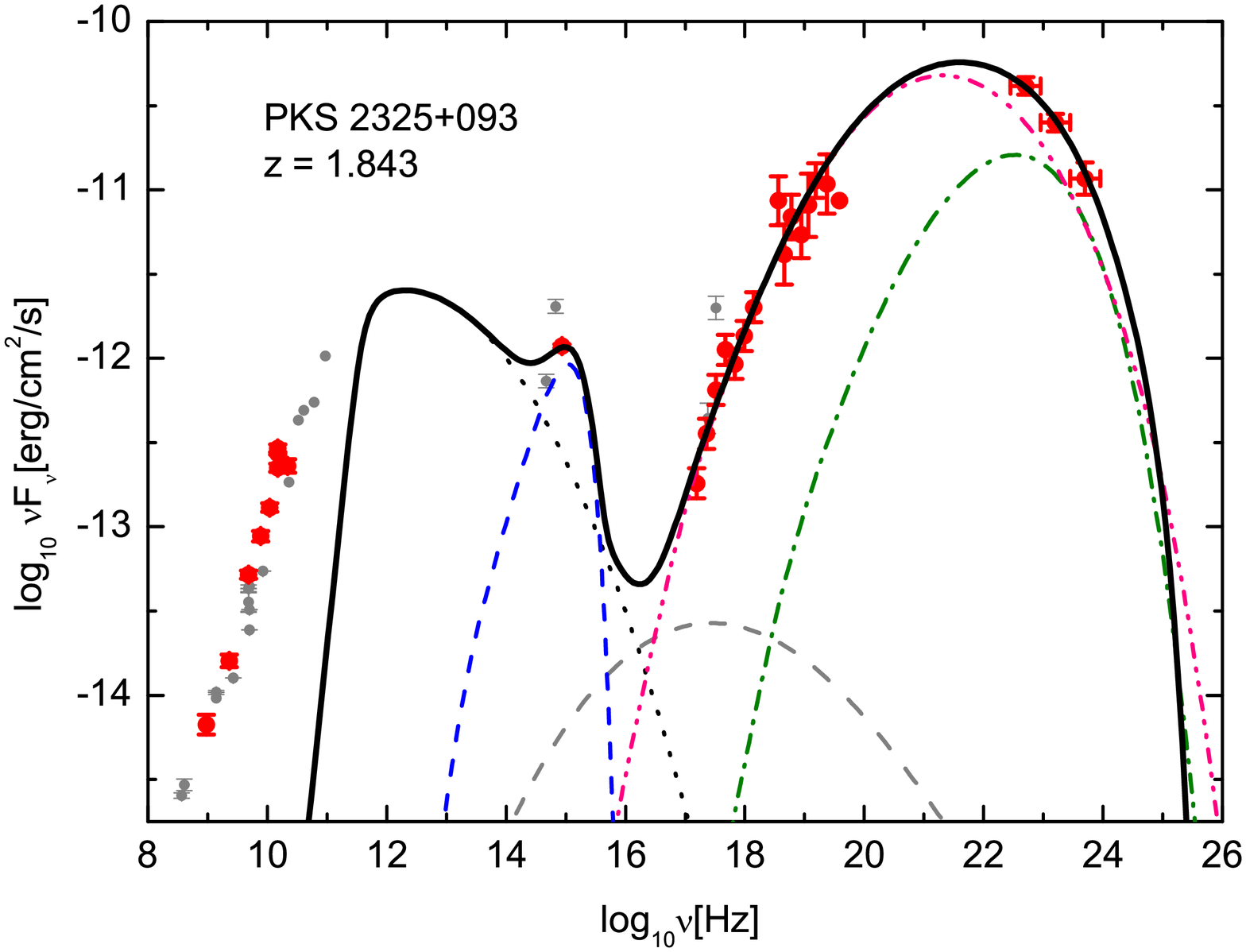}
 \includegraphics[width=7.0cm]{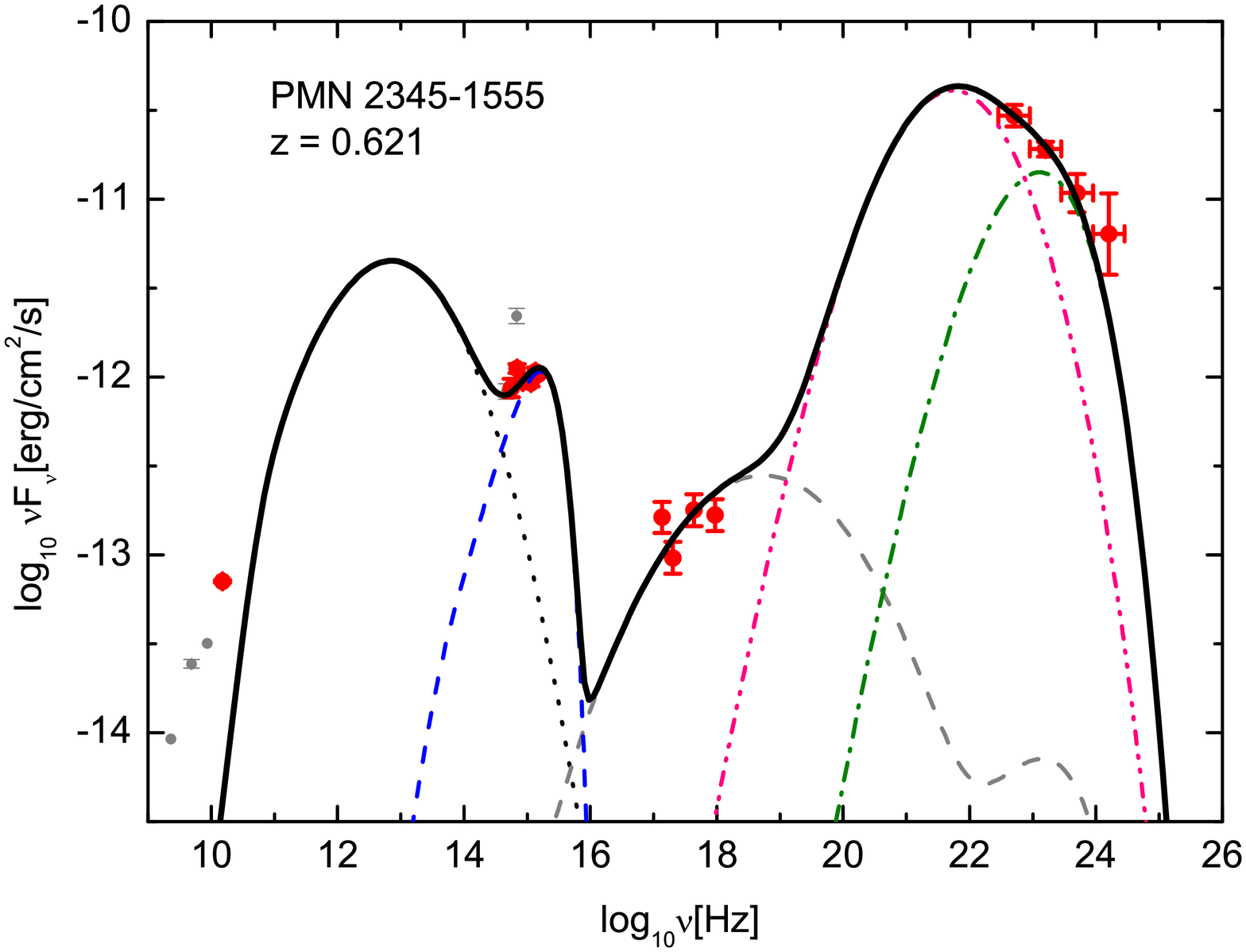}\\
 \caption{Same as Figure~\ref{figsed1}, but for different sources.}\label{figsed3}
\end{figure*}

\acknowledgments
This work is partially supported by the National Science Foundation of China (grants U1531131, 11433004) and the Top Talents Program of Yunnan Province.
We acknowledges funding support by Key Laboratory of Astroparticle Physics of Yunnan Province (No. 2016DG006).

\nocite{*}

\end{document}